\documentclass[conference]{IEEEtran}
\usepackage[normalem]{ulem}
\usepackage{booktabs} 
\usepackage{mathptmx} 
\usepackage{fancyhdr}
\usepackage{hyperref}
\usepackage{multirow}
\usepackage{graphicx}
\usepackage{caption}
\usepackage{pifont}
\usepackage{amsmath}
\usepackage{float}
\usepackage{color}
\usepackage{array}
\usepackage[flushleft]{threeparttable}
\usepackage[bottom]{footmisc}
\usepackage{listings}
\usepackage{datetime}
\usepackage{amsfonts}
\usepackage{algorithm}
\usepackage[noend]{algpseudocode}
\usepackage[figuresleft]{rotating}
\usepackage{subdepth}
\usepackage{subcaption}
\usepackage[table]{xcolor}
\usepackage{rotating}
\usepackage{adjustbox}

\newcommand{\bheading}[1]{{\vspace{2pt}\noindent{\textbf{#1}}\hspace{2pt}}}
\newcommand{\etal}{\textit{et al.}}

\ifCLASSINFOpdf
\else
\fi
\begin{document}
%
\title{CloudShield: Real-time Anomaly Detection in the Cloud}

\author{\IEEEauthorblockN{Zecheng He}
\IEEEauthorblockA{Princeton University\\
zechengh@princeton.edu}
\and
\IEEEauthorblockN{Ruby Lee}
\IEEEauthorblockA{Princeton University\\
rblee@princeton.edu}
}


%



\maketitle

\begin{abstract}
In cloud computing, it is desirable if suspicious activities can be detected by automatic anomaly detection systems. Although anomaly detection has been investigated in the past, it remains unsolved in cloud computing. Challenges are: characterizing the normal behavior of a cloud server, distinguishing between benign and malicious anomalies (attacks), and preventing alert fatigue due to false alarms.

We propose CloudShield, a practical and generalizable real-time anomaly and attack detection system for cloud computing. Cloudshield uses a general, pretrained deep learning model with different cloud workloads, to predict the normal behavior and provide real-time and continuous detection by examining the model reconstruction error distributions. Once an anomaly is detected, to reduce alert fatigue, CloudShield automatically distinguishes between benign programs, known attacks, and zero-day attacks, by examining the prediction error distributions. We evaluate the proposed CloudShield on representative cloud benchmarks. Our evaluation shows that CloudShield, using model pretraining, can apply to a wide scope of cloud workloads. Especially, we observe that CloudShield can detect the recently proposed speculative execution attacks, e.g., Spectre and Meltdown attacks, in milliseconds. Furthermore, we show that CloudShield accurately differentiates and prioritizes known attacks, and potential zero-day attacks, from benign programs. Thus, it significantly reduces false alarms by up to 99.0\%.
\end{abstract}


%

\newcommand{\csfigurepath}{figures}
\newif\ifisthesis
\isthesisfalse

\section{Introduction} \label{sec:intro}

The importance of cloud computing has grown significantly in the past years. Cloud customers can lease virtual machines from the cloud providers economically, sharing the physical resources provided by the cloud computing servers. Large cloud providers, like Amazon AWS \cite{amazonaws}, Google Cloud Platform \cite{googlecloud}, and Microsoft Azure \cite{microsoftazure}, have proliferated this trend.

There have been various attacks against cloud computing, especially on shared resources. For example, security-critical information, e.g., encryption keys, can be leaked by cache side-channel attacks. Previous work have revealed that many types of cache side-channel attacks can successfully obtain secret or private cryptographic keys \cite{bonneau2006cache, osvik2006cache, gullasch2011cache, yarom2014flush, zhang2014cross, gruss2015cache, irazoqui2015s, liu2015last, kayaalp2016high}. Recently, speculative execution attacks \cite{kocher2019spectre, lipp2018meltdown, kiriansky2018speculative} exploit performance optimization features of modern processors to breach the user-user, user-kernel or user-hypervisor isolation. Besides, zero-day attacks introduce challenges as they do not have known code nor known behavior.

Anomaly detection techniques are perhaps the only viable solution for detecting unknown zero-day attacks. By its nature, anomaly detection does not look for specifics of an attack but models the normal behavior of a system. Deviation from normal behavior indicates anomalies: either an attack or a benign anomaly.

However, existing anomaly detection systems in the cloud have challenges. First, a model of cloud server behavior is usually scenario-specific and is not easy to extend \cite{bossi2016system, asselin2016anomaly, garg2020abc}. Multiple models have to be built to cover various cloud workloads. Second, false alarms in anomaly detection systems are very common in practice. The large volume of false alarms overwhelms the security analysts and causes alert fatigue, potentially causing real attacks to be missed.

In this work, we investigate three questions. First: \textit{Can we make an anomaly detection system generalizable to different scenarios in cloud computing?} We hypothesize that the normal behavior of a cloud server, although different from workload to workload running on it, consists of a major predictable part, and a minor unpredictable part that follows a certain probability distribution. If we pre-train a general model to predict a cloud system's behavior, an anomaly can be detected by subtracting the predictable part from the original behavior markers and identifying the distribution of the remaining unpredictable part. To this end, we propose that the distribution of the unpredicted part denoted \textbf{Reconstruction Error Distribution (RED)}, can capture the characteristics of any cloud workload. Thus, we show that rather than deploying an individual model for each workload, a general pretrained predictor model is leveraged, and anomalies are identified by statistically comparing the REDs.

The second question we investigate is: \textit{How to select appropriate behavior markers to detect anomalous behavior in the cloud in real-time?} Quick detection of anomalies and attacks can prevent further damage. To support real-time anomaly detection in the cloud, we need an approach to select appropriate behavior markers that can be measured at high frequency and can reliably represent the system's behavior. To this end, we propose a principal component analysis (PCA)-based behavior marker selection method, and leverage the hardware performance counters, which are originally designed to monitor system performance and can be measured at high frequency, as exemplary markers to support real-time protection. 

The third question we explore is: \textit{How to deal with false-alarm fatigue?} In practice, the ``benign anomalous'' behavior of a cloud system is quite common. For example, a cloud server used for database applications may be scheduled a different task when its workload is low. The missing piece in the past anomaly detection is the ability to correctly recognize the new tasks as benign. Otherwise, a large number of false alarms are raised, causing the system to be no longer usable. In this work, we refine each detected anomaly with the identification of benign anomalies and known attacks as a second step. This can significantly alleviate the false alarm problem in anomaly detection.


Section \ref{sec:bg} describes the background. Section \ref{sec:threat} presents the threat model. Section \ref{sec:cs} discusses key challenges for anomaly detection in cloud computing. Section \ref{sec:impl} describes our CloudShield methodology and Section \ref{sec:exp} evaluates our design.

\section{Background} \label{sec:bg}

\subsection{Attacks in Cloud Computing}

\ifisthesis
There have been many attacks on cloud computing. We focus on the rapidly growing and representative class of software attacks on shared hardware resources in cloud servers in this chapter.
\else
There have been many attacks on cloud computing. We focus on the rapidly growing and representative class of software attacks on shared hardware resources in cloud servers.
\fi
Two main types are speculative execution attacks and cache-based side-channel attacks, which we use as example attacks in the evaluation of our anomaly detection system. We also include software attacks, e.g., buffer overflow, in our evaluation. Our system is not tailored at all to defeat these attacks, and the goal of our system is to detect even zero-day attacks, which are attacks that have never been seen before.

\subsubsection{Speculative Execution Attacks}

Since their first appearance in January 2018, speculative execution attacks \cite{kocher2019spectre, lipp2018meltdown, meltdown3a, kiriansky2018speculative, Spectrev4, koruyeh2018spectre, weisse2018foreshadow} have bombarded the world, with new variants continuously popping up. These attacks can leak the entire memory and break the software isolation provided by different virtual machines in the cloud, different virtual address spaces, and even by secure enclaves provided by SGX \cite{weisse2018foreshadow, brunellaforeshadow}. Speculative attacks misuse the hardware performance optimization features in modern processors, e.g., Out-of-Order (OoO) execution, speculative execution, hardware prediction, and caching. These attacks allow transient instructions to execute, illegally access a secret, and change the microarchitectural state based on the secret \cite{he2021new}. When the transient instructions abort, architectural changes are discarded but the microarchitectural changes, e.g., cache state changes, remain. This leaves an opportunity for obtaining the secret by monitoring the microarchitectural state.

\subsubsection{Cache Side-channel Attacks}

Cache-based side-channel attacks are timing attacks that have traditionally been used to leak the secret key of symmetric-key ciphers or the private key of public-key ciphers, thus nullifying any security provided by such cryptographic protections \cite{he2017secure}. They can be classified based on cache ``hit'' or ``miss'', ``access'' or ``operation''. The access-based attacks leverage the difference in timing between a hit and a miss to indicate a ``1'' or ``0'' based on single memory access, while the operation-based attacks leverage the time difference for a whole encryption operation. These attacks target different levels of the cache hierarchy, e.g., L1 cache and last-level cache (LLC).

Two representative cache side-channel attacks are the flush-reload attack and prime-probe attack. In the flush-reload attack, the initial state of a shared cacheline is set to absent by a \textit{clflush} instruction. After waiting for a while for the victim to execute, if the victim did use the cacheline, the attacker will find the cacheline present (indicated by a fast cache hit). A variant of the flush-reload attack, i.e., the flush-flush attack \cite{gruss2016flush}, exploits the early abort if the cacheline to be flushed is not in the cache.

In the prime-probe attack, the attacker first loads his data to fill the cache. After waiting for the victim to execute, the attacker checks (probe) if his cache lines are now absent, i.e., a slow cache miss, because the victim has brought in his data that evicted the attacker's cache lines.

\ifisthesis
\bheading{Hit and access-based attacks, e.g., the Flush-Reload attack \cite{yarom2014flush}.} The initial state of a shared cacheline is set to absent by a \textit{clflush} \footnote{Evict the cacheline at the specified address.} instruction by the attacker program. Then, if the victim program does not use the shared cacheline, the check (Reload) by the attacker program, after waiting for a while for the victim to execute, will still find the cacheline absent (indicated by a slow cache miss). If the victim did use the cacheline, then the attacker will find the cacheline present (indicated by a fast cache hit). A variant of the Flush-Reload attack, i.e., the Flush-Flush attack \cite{gruss2016flush}, exploits the early abort if the cacheline to be flushed is not in the cache.
\else
\fi




\subsubsection{Buffer Overflow Attacks}

A buffer overflow attack \cite{wang2008sigfree} occurs when the written data exceeds the size of an allocated buffer. Buffer overflow attacks can be exploited by an attacker to insert code and data. A buffer overflow attack is usually triggered by malformed input to write executable code or malicious data to a destination that exceeds the size of the buffer. If the malicious code or wrong data is used in the program, erratic program behavior would occur, e.g., system crash, incorrect results, or incorrect privilege escalation.

\subsection{Hardware Performance Counters}

Hardware performance counters (HPCs) are special registers that record hardware events. HPCs are widely available in commodity processors, including Intel, ARM, AMD, and PowerPC. Processors have been equipped with a Performance Monitor Unit (PMU) to manage the recording of hardware events. HPCs measure hardware events like the number of cache references, the number of instructions executed, and the number of branch mis-predictions; they also measure system events, like the number of page faults and the number of context switches.
\ifisthesis
There are two working modes for collecting HPC measurements. The first mode is \textit{counting}, where the occurrence of events are aggregated during program execution. The second mode is \textit{sampling}, where an interrupt is introduced when an event exceeds the pre-defined threshold. Counting-based HPC measurement provides more flexibility of measurement, e.g., collecting HPC readings during a fixed time or number of cycles. Therefore, we use the counting-based measurement for CloudShield in our design.
\fi

Although the HPCs were designed for system performance monitoring and software debugging, previous work have also shown the feasibility of using hardware performance counters in security, e.g., detecting malware \cite{ozsoy2016hardware, demme2013feasibility}, firmware-modification \cite{wang2015confirm} and kernel root-kits \cite{wang2014detecting}. Zhou \etal \cite{zhou2018hardware} and Das \etal \cite{das2019sok} cautioned using HPCs for security. Unlike these existing work, CloudShield leverages the reconstruction error distribution of HPCs, rather than directly using the noisy HPCs for anomaly detection. \cite{ozsoy2016hardware}, \cite{yin2017deep} and \cite{patel2017analyzing} exploited hardware performance counters and supervised neural networks for malware and intrusion detection, respectively. A major drawback of using supervised deep learning for attack detection is that they require attack examples for training. Thus the zero-day attacks which are not seen in the training set can not be detected at runtime.

\section{Threat Model}\label{sec:threat}


The target system is a cloud-based Infrastructure-as-a-Service (IaaS) system, where programs share hardware resources. The programs running on the IaaS platform may interfere with each other. As is commonly done, important and frequently used cloud services are scheduled one main task per machine, or per processor core, e.g., machine learning training, database query, MapReduce, or being used as a web or stream server. New tasks can be scheduled on the same core if the workload of the main task is low.

Our threat model covers attacks that breach the confidentiality and integrity of the cloud computing system. Notably, the side-channel attacks and the recently proposed speculative execution attacks are considered in this threat model. We assume the attacker can launch attack programs in the cloud. We assume that an attack program can hide by switching between running and sleeping.

Our threat model particularly includes zero-day attacks. Unlike signature-based attack detection, we do not make particular assumptions about the attacks. We assume that there is no prior knowledge of attack code and the way the adversary interferes with the system.

Furthermore, once an anomaly is detected, we explicitly consider reducing false alarms caused by other benign programs that concurrently run. These benign programs need to be distinguished from attacks, otherwise, they can cause a large number of false alarms. Consequently, cyber analysts can be overwhelmed by false alarms and miss real attacks, making the detection system ineffective in practice. Therefore, discriminating benign programs, known attacks, and zero-day attacks is an important component in this work.

From the system design perspective, the rapidity and generalizability of the detection system are important. First, real-time anomaly detection is crucial, because an attack, e.g., the side-channel attacks and the speculative execution attacks, can quickly achieve their goal of leaking secrets in less than a minute. Quick detection and mitigation of the attacks prevent more damage to the system. We assume the hardware performance counters (HPCs) can be measured in the system. This assumption is reasonable because most of the modern processors used in cloud computing servers have been equipped with HPCs, and the mainstream operating systems support collecting HPC measurements. Second, generalizability plays an essential role in our design. We would like as few models as possible, ideally, a single model, to cover various cloud scenarios and workloads with minimum changes. This reduces the cost of switching models between workload changes.

\section{Cloudshield Challenges} \label{sec:cs}

We first identify three challenges of anomaly detection in the cloud, and how they can be handled:
\begin{enumerate}
    \item How to model the different cloud workloads?
    \item How to select appropriate behavior markers?
    \item Anomaly detection can cause false-alarm fatigue. How to deal with the false alarms of anomaly detection in the cloud?
\end{enumerate}

\subsection{How to Model Different Cloud Workloads?} \label{sec:cs:model}

Our intuition of modeling cloud workloads, which may vary a lot in their functionalities, scales, and required resources, is that \textbf{\textit{the behavior of a cloud server running a common cloud workload can be decomposed into two parts: a major predictable component and a minor unpredictable component. The predictable component can be predicted by a pre-trained model. The unpredictable component follows an unknown but fixed distribution.}} We will validate this hypothesis in Section \ref{sec:exp}.

With this assumption, rather than an individual model for each workload, we can pre-train a general program behavior predictor model $M$ for the predictable component, and subtract the prediction from the observed measurement of the system. The distribution of the remaining unpredictable component, i.e., the reconstruction error distribution (RED), can reveal the normal behavior from abnormal behavior. We leverage RED as the key to anomaly detection. Stealthy attacks can be subtle and hide within normal measurements. However, subtracting the major predictable component of the measurements from the total observed measurements amplifies the anomalous behavior and provides a robust way of detecting sneaky anomalies.

\begin{figure}[h]
    \centering
    \includegraphics[width=0.5\linewidth]{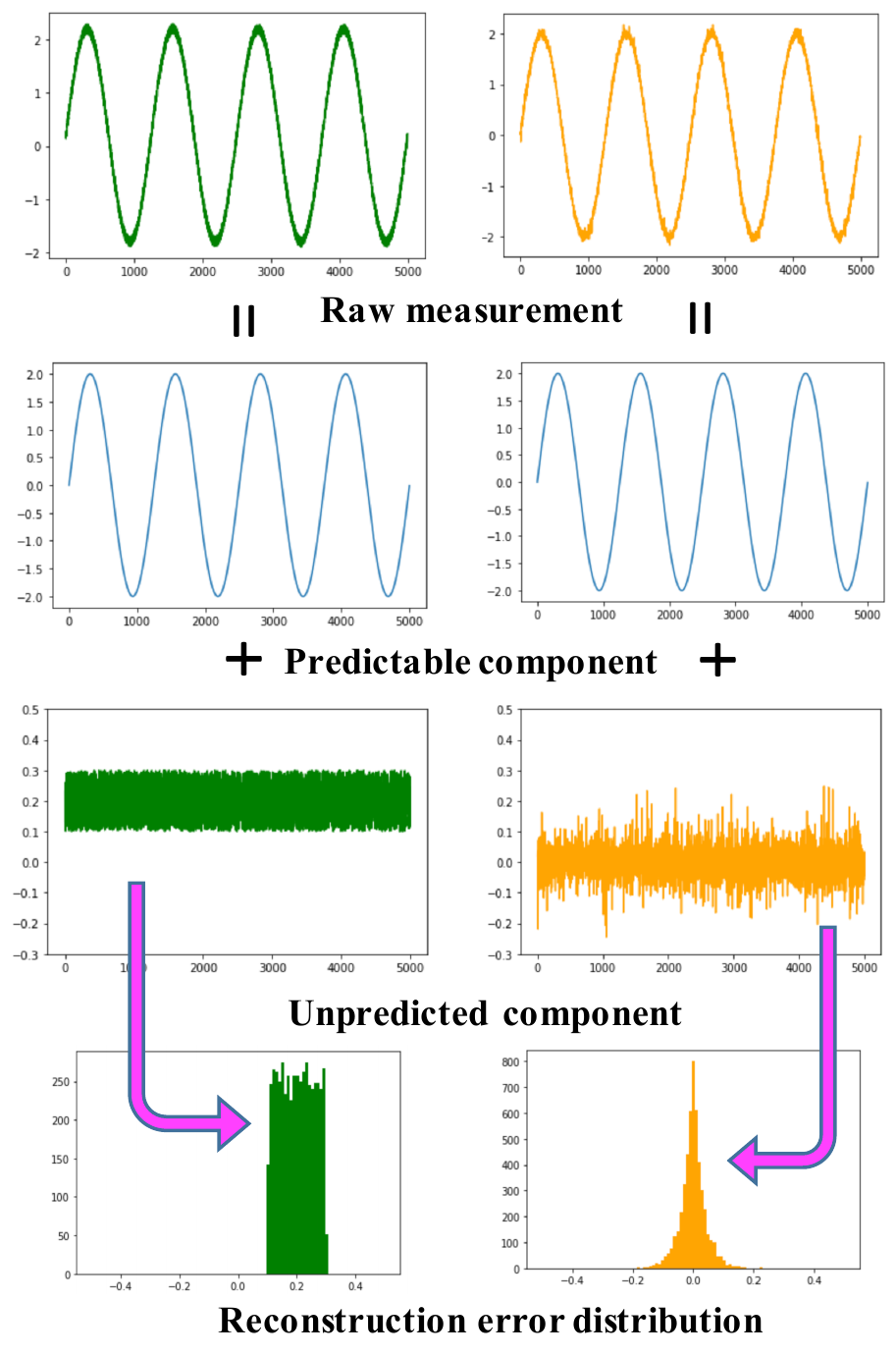}
    \caption{A running example of the reconstruction error distribution. Our intuition is that the behavior of a system can be decomposed into two parts: a major predictable component and a minor unpredictable component. If we can separate the predictable from the unpredictable component using a prediction model, the difference between normal and anomaly is more clearly revealed.} \label{fig:running-example}
\end{figure}

We present a running example to illustrate this idea in Figure \ref{fig:running-example}. Two sine curves plus subtle perturbations are shown in the top row, marked green and yellow, respectively. By looking at only these two raw measurements, one may not be able to tell the difference. We then subtract the predictable signal (the blue sine curves in the second row) to get the remaining part in the third row and examine the distribution of this remaining unpredictable part (bottom row). It shows that the probability distribution of the remaining part, which we denote the reconstruction error distribution (RED) from a prediction model, amplifies the difference.

\subsection{How to Select Appropriate Behavior Markers?} \label{sec:cs:behaviormarkers}

Modern processors usually provide counts of various events that can be used as behavior markers. Monitoring all of them is inefficient, if not impossible. Therefore, we need a method to choose the appropriate behavior markers from all possible markers that can represent the normal behavior of a system.

Our key idea for selecting behavior markers is to quantify the relative importance of the selected events representing the normal behavior of a system. Given the set of all possible behavior markers $b_1, b_2...b_n$, we can define a metric $f$ to evaluate the relative contribution of a marker in representing the normal behavior of the system. Then, the behavior markers are sorted in descending order according to $f(b)$. The markers that exceed a certain threshold of importance are selected as candidate markers. In our implementation, we define $f$ based on principal component analysis (PCA). Other metrics can also be leveraged to automatically select behavior markers.

\subsection{How to Distinguish Benign Anomalies and Malicious Attacks?} \label{sec:cs:benign}

The ultimate goal of CloudShield is to detect attacks, i.e., malicious anomalies. Once an anomaly is detected, the next step is to determine if it is a benign anomaly or a malicious attack. Without loss of generality, we simplify the discussion by making the assumption that a processor core runs one cloud workload, e.g., a stream server or a web server. A malicious anomaly can be a known attack or a zero-day attack. A benign anomaly can be benign programs that run concurrently with the cloud workload, where their interference could potentially cause false alarms. It could also be a stealthy attack that looks like a benign program. Note that the key difference between a cloud workload and a benign program is that the cloud workload, as is commonly done, is the one main task per cloud server, or per processor core, while benign programs are relatively small programs that can be scheduled on the same core if the workload of the main task (cloud workload) is low.

While anomaly detection systems typically fall short of detecting benign versus malicious anomalies, Cloudshield can detect not just anomalies, but also the subset of anomalies that are attacks. Specifically, CloudShield builds two detectors, one is to identify known benign programs, and the other is to identify attacks. Also, while actual attack detection tends to be very domain-specific, our new contribution is to show that it is possible to use a general framework based on a pre-trained model to do attack detection. We are even able to detect stealthy attacks and potential zero-day attacks.





\section{CloudShield} \label{sec:impl}

\subsection{Overview}

\begin{figure*}[h]
    \centering
    \includegraphics[width=0.8\linewidth]{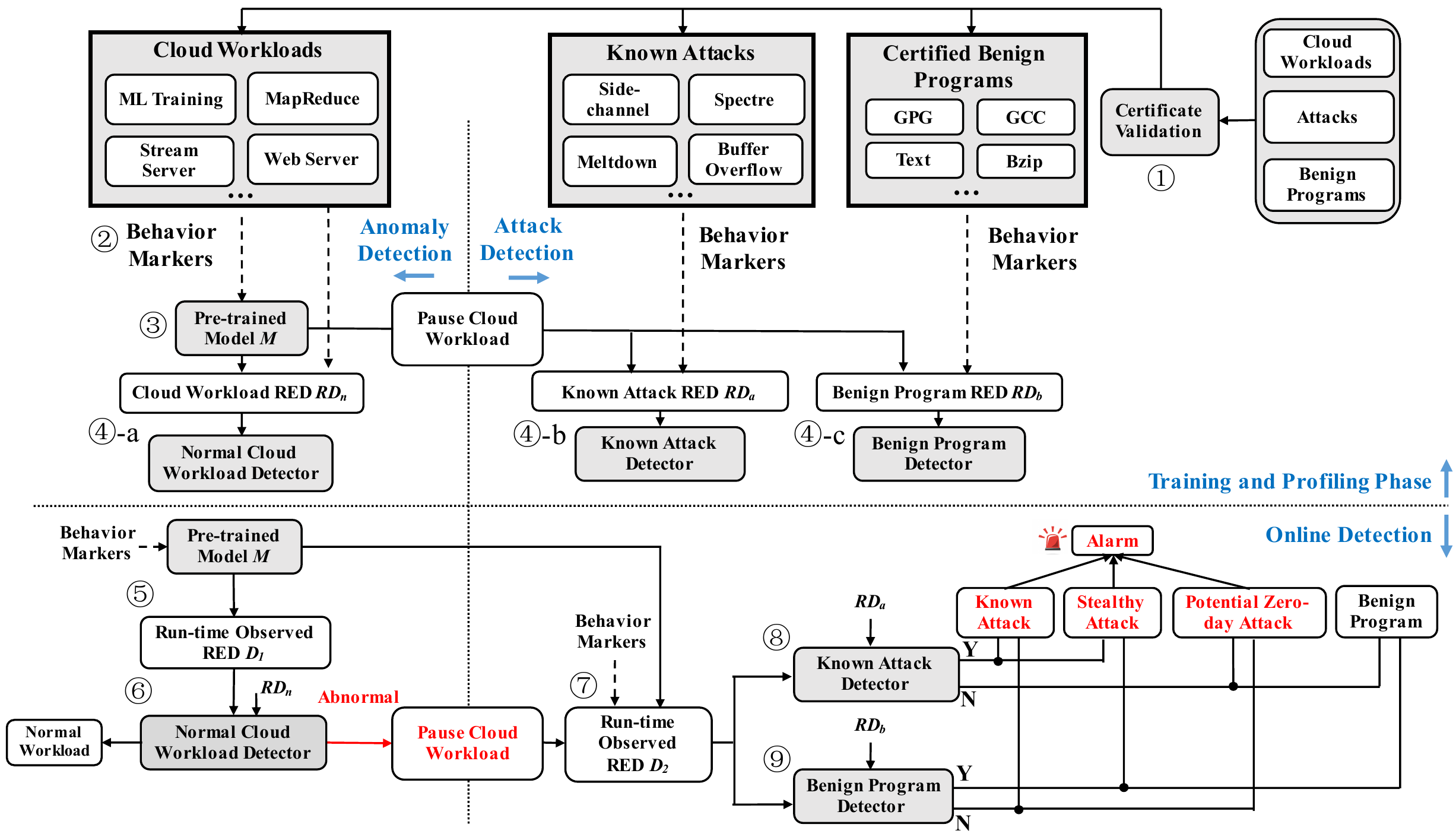}
    \caption{CloudShield methodology for anomaly and attack detection.}\label{fig:overview}
\end{figure*}

We show an overview of CloudShield in Figure \ref{fig:overview}. There are three phases for learning and detecting anomalies and attacks in the cloud: \textit{1) offline training and profiling}, \textit{2) online anomaly detection and mitigation} and \textit{3) online attack versus benign program detection}.

The \textit{offline training and profiling phase} consists of four steps:

\ding{172} constructing three sets of programs: normal cloud workloads, known attacks, and certified benign programs. A \emph{Certificate Validation Module} is responsible for verifying the certificates of the workload and benign programs. The certificates are generated by trusted entities, e.g., companies that create these programs, and organizations or labs that verify the correctness and security of the programs. The certificate must contain the hash of the program binary and the public key signature of the trusted entity.

\ding{173} Executing the workloads and programs in an offline clean environment and collecting their behavior markers. A \emph{Program Behavior Collection Module} is designed for this.

\ding{174} Training a default program-behavior predictor model $M$ in a \emph{Training Module}.

\ding{175} Calculating the corresponding REDs $RD_n$, $RD_a$, and $RD_b$ as the reference Reconstruction Error Distributions (REDs) for normal cloud workloads, known attacks, and benign programs, respectively. We use the distribution $RD_n$ as the normal behavior of the processor core running a cloud workload, while $RD_a$ and $RD_b$ are used to further distinguish between known attacks and benign programs when an anomaly is identified. Note that the cloud workload needs to be paused before collecting HPCs and calculating REDs for attacks and benign programs. The normal cloud workload detector, known attack detector, and benign program detector are also computed at this time.

The \textit{online anomaly detection and mitigation phase} has two steps:

\ding{176} An \emph{Online Detection Module} collects runtime behavior markers of each processor core in a cloud server from the Performance Monitor Unit (PMU) in the host OS. These markers are input into the pre-trained model $M$ for the inference phase, to generate the run-time observed RED $D_1$.

\ding{177} Comparing the run-time RED $D_1$ and the reference RED $RD_n$. If $D_1$ does not follow the distribution of normal cloud workloads $RD_n$, an anomaly is detected and the cloud workload is paused to avoid further security breaches.

Once an anomaly is detected, the \textit{online attack versus benign program detection phase} (step2 detection) is performed to distinguish benign programs from known attacks. This phase has three steps:

\ding{178} Collecting behavior markers when the cloud workload is no longer running. This step is necessary to eliminate the interference from the cloud workload, which is usually heavy, and increase the detection accuracy. The new measurements are inferred through the pre-trained model $M$ and new RED $D_2$ is gathered.

\ding{179} Comparing $D_2$ to the distribution of known attacks $RD_a$ to identify if the anomaly is caused by a known attack.

\ding{180} Comparing $D_2$ to the distribution of certified benign programs $RD_b$ to identify if the anomalous behavior is a false alarm. Note that the steps \ding{179} and \ding{180} can be performed in parallel. As a complementary component, the cloud provider can confirm that benign programs are scheduled on this machine.

In the above discussion, we have assumed that a single pre-trained model of normal cloud workloads is sufficient, and that different known attacks can be detected with a single known attack detector, and that all benign programs added to a cloud workload can be identified with a single benign program detector. This significantly simplifies the implementation of CloudShield, and we will show that this results in excellent anomaly and attack detection in practice. More cloud workloads, attacks, and benign programs can always be added to the three sets of programs to retrain the model $M$ and the three detectors.

The CloudShield implementation consists of four modules: a certificate validation module, a program behavior collection module, a training module, and an online detection module. The servers can share a set of the first three modules, as they are used during training phase. Only the last module needs to run on each cloud server.

\subsection{Pre-training Program Behavior Predictor} \label{sec:cs:pretrain}

\bheading{Feature selection.} Modern processors usually provide various events to be monitored by using hardware performance counters. However, due to the limited number of hardware registers in the PMU, only a few of them can be monitored at the same time. While round-robin scheduling of HPC measurements is feasible, it increases overhead. Therefore, it is important to select the appropriate events from all possible events as behavior markers. We propose a principal component analysis (PCA) based selection method to help determine the events to monitor. Our key idea is the selected events should be important to represent normal behavior.

Specifically, the principle component $PCA_{1}$ can be represented as a linear combination of all features. The coefficient of the corresponding HPC measurement represents the contribution of that feature in the principal component. Formally,
\begin{align}
    PCA_{1} &= ||x^Tw||^2 \\
          &= \sum_i |w_i|^2 x_{i}^2
\end{align}
where $x=(x_1, x_2...x_n)$ is an HPC reading of $n$ events. $|w_i|$ is the coefficient of $x_i$ in the first principal component. It represents the importance of event $x_i$ in the first principal component.

\ifisthesis
We collect 34 HPC events (Table \ref{tab:cs:hpc-list}) from five representative cloud benchmarks, i.e., ML training (PyTorch), stream server (FFserver), database server (Mysql), web server (Nginx), and Hadoop MapReduce. We collect the event measurements for an entire processor core, to provide system-level monitoring, rather than just monitor a specific process or thread. We then perform the PCA analysis on each benchmark. We present the top-10 events for each benchmark in Table \ref{tab:cs:feature-selection-per-task}. We also mark the events that achieve top-10 in all five benchmarks in bold font. We observe that although the benchmarks are different, they show consistency in the events' importance. Only seven common events in top-10 of all five benchmarks: number of instructions, cycles, loads, stalls during retirement, stalls during issue, DTLB reads, and DPU reads. The number of stores and DTLB writes occur in four of five benchmarks. Other top-10 events are the number of L1D read misses, L1I read misses, branches and context switches. The thirteen events in Table \ref{tab:cs:feature-eta} correspond to all events that occured as one of the top-10 events for any of the benchmark.
\else
We collect 34 HPC events (Table \ref{tab:cs:hpc-list} in Appendix) from five representative cloud benchmarks, i.e., ML training (PyTorch), stream server (FFserver), database server (Mysql), web server (Nginx), and Hadoop MapReduce. We collect the event measurements for an entire processor core, to provide system-level monitoring, rather than just monitor a specific process or thread. We observe that although the benchmarks are different, they show consistency in the events' importance.
\fi

We use $\eta_i = \frac{|w_i|}{\sum_j |w_j|}$ as the importance of the corresponding event for a workload. We average $\eta$ over the five representative benchmarks as the final importance score $\overline{\eta}$ of the corresponding event. We show the features with $\overline{\eta} \ge 1\%$ in Table \ref{tab:cs:feature-eta}. We use the thirteen selected events throughout the experiments. In fact, these are also the thirteen distinct events in the top-10 events for the five cloud workloads \ref{tab:cs:hpc-list}.

\begin{table}[h]
\centering
\caption{HPC features with $\overline{\eta} \ge 1\%$.} \label{tab:cs:feature-eta}
\resizebox{\linewidth}{!}{
\begin{tabular}{|c|c|c|c|c|c|}
\hline
\textbf{Rank} & \textbf{Event}          & \textbf{$\overline{\eta}$} & \textbf{Rank} & \textbf{Event} & \textbf{$\overline{\eta}$} \\ \hline
\textbf{1}    & Instruction             & 0.267                    & \textbf{8}    & BPU read       & 0.030                    \\ \hline
\textbf{2}    & Stall during issue      & 0.189                    & \textbf{9}    & DTLB write     & 0.025                    \\ \hline
\textbf{3}    & Stall during retirement & 0.178                    & \textbf{10}   & Branch         & 0.023                    \\ \hline
\textbf{4}    & Cycles                  & 0.106                    & \textbf{11}   & L1D read miss  & 0.020                    \\ \hline
\textbf{5}    & Load                    & 0.067                    & \textbf{12}   & L1I read miss  & 0.018                    \\ \hline
\textbf{6}    & DTLB read               & 0.043                    & \textbf{13}   & Context switch & 0.015                    \\ \hline
\textbf{7}    & Store                   & 0.037                    &               &                &                          \\ \hline
\end{tabular}
}
\end{table}

\bheading{Model selection.}
\ifisthesis
Deep learning has shown great power in many domains, e.g., computer vision \cite{lecun1998gradient, krizhevsky2012imagenet, he2015delving, russakovsky2015imagenet} and natural language processing \cite{devlin2018bert, lan2019albert}. It inspires us to investigate the feasibility of using deep learning to model the normal behavior of a system.
\fi
Recurrent Neural Network (RNN) and its variant, Long Short-Term Memory (LSTM), have become the popular model for sequential data. To balance the model complexity and its prediction power, in the proof-of-concept implementation, we start from a single-cell LSTM as the behavioral model of the system. An LSTM cell has three gates that control information flow: the forget gate, the input gate, and the output gate. LSTM automatically determines what information to ``remember'' and ``forget''.

Alternative models, e.g., Gated Recurrent Units (GRUs) \cite{GRU} and BERT \cite{BERT}, can also be used as behavioral models of the system. As the main focus of this work is not to find the best model, but to show the feasibility of using RED of HPCs to detect anomalies in the cloud system, without loss of generality, we just show that LSTM models are enough for this anomaly detection.

\bheading{Model training.} Our goal is to train a model that can capture the predictable component of the behavior of a program. The program behavior markers $\{S_{i}\}_{i=1}^N$ (in our case HPC events of cloud workloads), are obtained from a clean environment. $N$ is the total number of time frames collecting HPCs. In our experiments, each behavior measurement $S_i^t$ is a vector consisting of the thirteen monitored hardware events. At time $t$, the deep learning model is trained to predict $S_i^{t+1}$ using behavior history $[S_i^1, ..., \ S_i^{t}]$. Intuitively, since $\{S_{i}\}_{i=1}^N$ are normal behavior markers collected in the clean environment, the loss penalizes the incorrect prediction of normal behavior. We train this model to minimize the loss function with Stochastic Gradient Descent (SGD).


\subsection{RED Profiling} \label{sec:cs:profiling}

\bheading{RED generation of cloud workloads.} We generate a profile of the normal cloud workloads in terms of reconstruction error distribution (RED), illustrated as $RD_n$ in Figure \ref{fig:overview}. First,
\ifisthesis
as in Chapter \ref{ch:power-grid} Section \ref{sec:power-grid:power-method:offline},
\fi
reference sequences of the behavior measurement, $R=[R^1,..., R^{T'}]$, are collected in a clean environment. For this cloud server setting, each time frame $R^i$ is a vector of thirteen dimensions (the number of monitored events) in our experiment.
\ifisthesis
\footnote{Different from Chapter \ref{ch:power-grid} Section \ref{sec:power-grid:power-method:offline} where the behavior markers are measured for each thread (23x4), here we monitor thirteen behaivor markers per core on the system level.}
\fi
Second, at time frame $t$, we use the trained model to predict time frame $t+1$ using the corresponding history behavior. We denote the prediction as $P^{t+1}$. The reconstruction error is defined as:
\begin{align}
E(t)=R^{t+1}-P^{t+1} \label{equ:cs:reconstructionerror}
\end{align}
\ifisthesis
Note that, different from Eq. \ref{equ:power-grid:reconstructionerror} where we calculate the magnitude of the errors because KS test only works for the one-dimensional scenario, here each reconstruction error sample $E(t)$ is a vector of dimension $n$,
\else
Each reconstruction error sample $E(t)$ is a vector of dimension $n$,
\fi
where $n$ is the number of monitored events. We gather the prediction errors of each cloud workload and define the overall distribution of \{E(1), E(2), E(3)...\} from \textit{all} workloads as $RD_n$.


\bheading{KDE profiling of cloud workload.}
\ifisthesis
One limitation of the KS test in Chapter \ref{ch:power-grid} is that it only applies to an one-dimensional scenario. \footnote{In Eq.\ref{equ:power-grid:reconstructionerror}, high-dimensional reconstruction errors are mapped to one-dimension by calculating magnitude.} Therefore, we use Kernel Density Estimation (KDE),
\else
We use Kernel Density Estimation (KDE),
\fi
a non-parametric estimation approach that better handles high-dimensional data, to profile the high-dimensional distribution of reconstruction errors from reference samples, denoted \ding{175}-a in Figure \ref{fig:overview}. We use non-parametric estimation because the formula of the RED of normal workloads is unknown, and its formula can be too complex to assume. KDE represents the distribution from elementary kernels. It assumes a small high probability area (Gaussian in our implementation) within a bandwidth around the observed samples, and sums them up as the probability distribution. Formally, KDE is defined as:
{\small
\begin{align}
    \hat{f}(x) = \frac{1}{nb} \sum_{i=1}^{n} K(\frac{x-x_i}{b}) \label{equ:kde}
\end{align}
}
where $\hat{f}(x)$ is the estimated probability density. $K(\cdot)$ is a kernel function, whose value drops rapidly outside a bandwidth $b$. $x_i$s are the samples from the distribution, i.e., E(t) in our case. $n$ is the total number of samples. 

\begin{figure}[h]
    \centering
    \includegraphics[width=\linewidth]{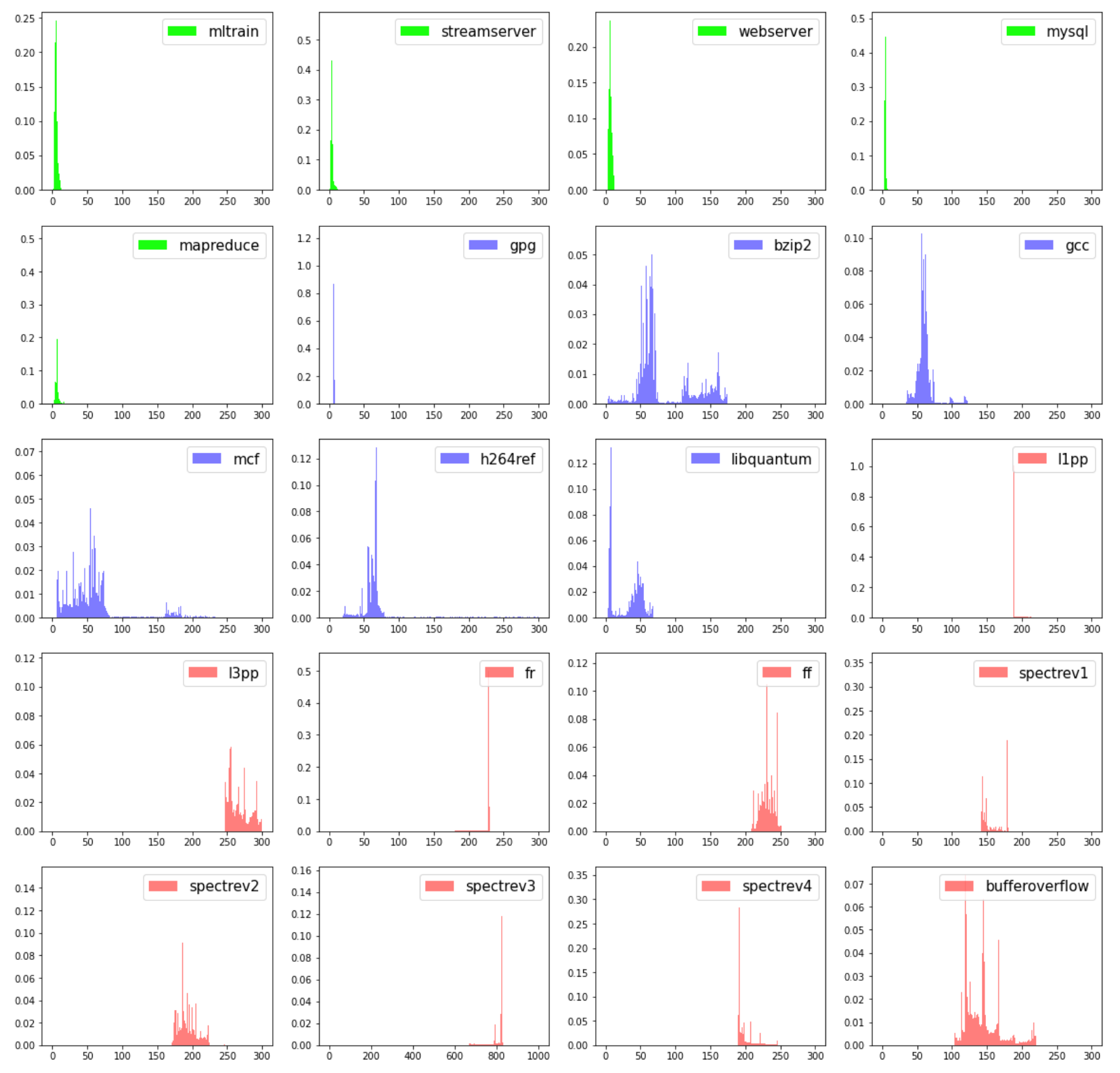}
    \caption{Reconstruction error distribution (RED) of normal cloud workloads (first five in green), benign programs (next six in blue) and attacks (last nine in red). RED of all attacks are different from the normal behavior of the cloud workload. The x-axis is the magnitude of RED, and y-axis is the probability density of that RED magnitude.} \label{fig:red-examples}
\end{figure}

In Figure \ref{fig:red-examples}, we show examples of reconstruction error distribution (RED) of normal cloud workloads (first five in green), benign programs (next six in blue), and attacks (last nine in red). To illustrate the high-dimensional distribution, we calculate the magnitude of REDs in Eq. \ref{equ:cs:reconstructionerror} and observe that the normal cloud workloads, in general, have the smallest REDs (distributions to the left). Figure \ref{fig:red-examples} shows clear difference between the cloud workloads, the benign programs, and the attacks. The cloud workloads have the smallest REDs (leftmost). The REDs of different benign programs are distinct, showing that RED can be used to distinguish benign programs. Most of the benign programs have larger REDs than cloud workloads, except the gpg-rsa program whose RED is similar to the cloud workloads. Moreover, the REDs of all evaluated attacks are to the right side, meaning larger reconstruction errors than cloud workloads and benign programs. We observe that one spectre attack (spectre v3) induces significantly larger RED than the other workloads, benign programs, and attacks. This shows that the spectre v3 attack's behavior is unique compared to the other attacks.

\bheading{Profiling for benign programs and known attacks.} Similarly, we profile the RED of the benign programs and known attacks. We collect their behavior data in a clean execution environment from the Program Behavior Collection Module. Interestingly, we observe that it is not necessary to train another program behavior predictor model for benign programs and attacks. The pre-trained one on cloud workloads can be reused to profile the benign programs and known attacks. We hypothesize that it is because pre-training on different workloads improves the generalizability of the model, by suppressing potential overfitting. At last, two KDE estimations are performed on the RED of known attacks and benign programs, shown as \ding{175}-b and \ding{175}-c in Figure \ref{fig:overview}, respectively.

\begin{figure}[h]
    \centering
    \includegraphics[width=0.5\linewidth]{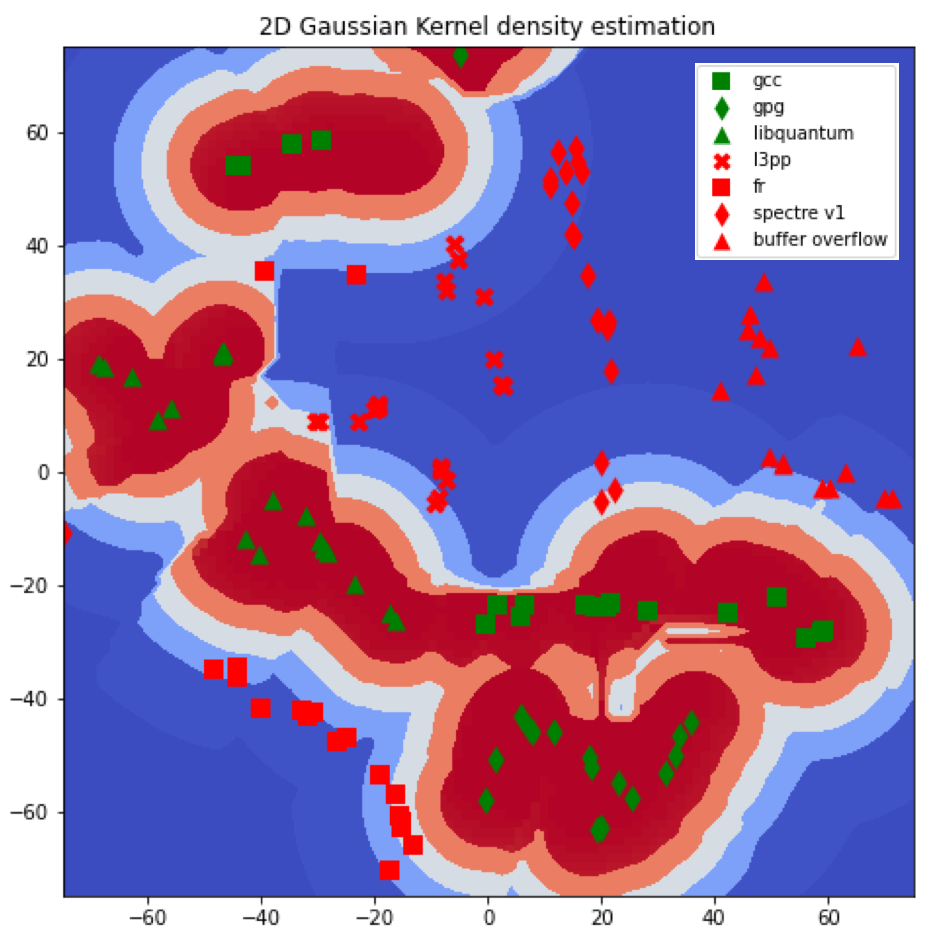}
    \caption{Illustration of kernel density estimation of benign programs. The high-density regions (likely to be benign programs) are colored red while the low-density areas (unlikely to be benign programs) are marked blue. We observe that benign programs (gcc, gpg, libquantum) are in the high-density area and attacks (l3pp, fr, spectre v1, and buffer overflow) are in the low-density areas.} \label{fig:kde-heatmap}
\end{figure}

We illustrate an example of kernel density estimation of benign programs in Figure \ref{fig:kde-heatmap}. To illustrate, we first use t-SNE \cite{hinton2002stochastic} to map the thirteen HPCs to a 2-D plane and build a KDE estimator of benign programs (gcc, gpg, and libquantum) using the REDs from the pre-trained model. The high-density regions (likely to be benign programs) are colored red while the low-density areas (unlikely to be benign programs) are colored blue. We plot three benign programs, i.e., gcc (green square), gpg (green diamond) and libquantum (green triangle) in Figure \ref{fig:kde-heatmap}. We also depict four attacks, i.e., l3pp (red cross), fr (red square), spectre v1 (red diamond), and buffer overflow (red triangle), in Figure \ref{fig:kde-heatmap} and observe that they are all in the low-density area, where the benign program detector can identify them as non-benign programs. Figure \ref{fig:kde-heatmap} explains why KDE works, specifically the benign programs form high-density clusters while the attacks are outside the clusters.

\subsection{Runtime Anomaly Detection and Mitigation} \label{sec:cs:ad}

The online detection module is responsible for detecting anomalies and distinguishing attacks and benign programs at runtime. A processor core's behavior, in terms of hardware event measurements, is dynamically monitored at runtime.

\bheading{Anomaly detection based on RED.} Similar to the offline profiling phase, the runtime gathered HPC sequences are sent through the pre-trained model (\ding{176} in Figure \ref{fig:overview}) to obtain the runtime observed RED $D_1$. The likelihood of the observed reconstruction error following the RED of normal cloud workloads ($RD_n$) is computed using the KDE normal workload detector ($\hat{f}(x)$ in Eq. \ref{equ:kde}) \footnote{Tree-based structures, e.g., KD tree, can be used to find the $x_i$s close to $x$ and accelerate the computation because the effect of $x_i$s outside the bandwidth $b$ is negligible.}. If the likelihood $\hat{f}(x)$ is lower than a pre-defined threshold, i.e., the prediction error does not follow the distribution of $RD_n$, an anomaly is detected.

Based on the results of the anomaly detection, different response actions can be taken. If no anomaly is detected, no further actions are required. Once an anomaly is detected, CloudShield triggers different responses (\ding{177} in Figure \ref{fig:overview}). First, the cloud workload running on the machine is temporarily paused to avoid further damage. This also eliminates the interference between the cloud workload and other tasks that concurrently run (attacks or benign programs). Second, access to the most security-critical data and resources is temporarily turned off. Attacks against data confidentiality, e.g., side-channels, can target these secret data. Thus, cutting access to the security-critical data prevents these data from being leaked out. Third, the known attack detector and benign program detector are woken up, to identify if the anomaly is malicious (an attack) or benign (a false alarm).
\ifisthesis
This can further reduce false-alarm fatigue in practice, as discussed in Section \ref{sec:cs:attack-and-benign}.
\else
This can further reduce false-alarm fatigue in practice, as discussed below.
\fi

\subsection{Distinguishing Benign Programs and Attacks} \label{sec:cs:attack-and-benign}

A detected anomaly can be caused by benign programs. Thus, CloudShield attempts to distinguish ``benign anomalies'' caused by benign programs versus real attacks. As discussed in Section \ref{sec:cs:ad}, the cloud workload is paused once an anomaly is detected (\ding{177} in Figure \ref{fig:overview}). Now the monitored core is possibly running attacks. Moreover, other benign programs (can be a victim program) that concurrently run with the attack may hide the attack and make identifying attacks even harder. We will show CloudShield can detect an attack in both scenarios, with and without benign programs running.

\bheading{Attacks and benign programs identification.} To distinguish the attacks and benign programs, firstly, hardware events' measurements are monitored through the PMU after the main cloud workload is switched off. Then the PMU sends the newly measured data (without cloud workload) to the same pre-trained program behavior predictor $M$ for inference. Similar to anomaly detection, we compute the RED $D_2$ in the form of Eq. \ref{equ:cs:reconstructionerror}. The KDE attack detector (\ding{175}-b) and the KDE benign program detector (\ding{175}-c) were loaded into the online detection module from the training module \footnote{Note that here we only need two KDE estimators, one for attacks and the other for benign programs, rather than an individual detector for each attack or benign program.}. The attack detector computes the likelihood of the observed prediction errors following the RED of known attacks ($RD_a$), using Eq. \ref{equ:kde}. If a high likelihood is observed, the attack detector reports an attack. Similarly, the benign program detector computes the likelihood of the observed prediction error following the RED of benign programs ($RD_b$). If a high likelihood is observed, the benign program detector reports a benign program.

Based on the decisions of the two detectors, we list the four possible final decisions in Table \ref{tab:cs:attack-benign}.

\begin{table}[h]
\centering
\caption{Benign program and attack decisions and responses.} \label{tab:cs:attack-benign}
\resizebox{\linewidth}{!}{
\begin{tabular}{|c|c|c|c|c|}
\hline
& \textbf{\begin{tabular}[c]{@{}c@{}}Known Attack  \\ Detector\end{tabular}} & \textbf{\begin{tabular}[c]{@{}c@{}}Benign Program  \\ Detector\end{tabular}} & \textbf{Decision} & \textbf{Response}       \\ \hline
\textbf{Case 1} & Y                               & Y                                 & Stealthy attack                         & Alarm (high priority)   \\ \hline
\textbf{Case 2} & Y                               & N                                 & Attack                                  & Alarm (high priority)   \\ \hline
\textbf{Case 3} & N                               & Y                                 & Benign program                          & Resume cloud workload   \\ \hline
\textbf{Case 4} & N                               & N                                 & Zero-day attack or new benign programs  & Alarm (medium priority) \\ \hline
\end{tabular}
}
\end{table}

\bheading{Case 1:} The attack detector recognizes it as a known attack, and the benign program detector recognizes it as a benign program. In this case, CloudShield reports it as a stealthy attack where the attack program hides by mimicking the behavior of a benign program. Another possible scenario of this case is that a benign program, which could be a victim program, is concurrently running with the attack program. We will show in the experiments that attacks can still be detected even when they run together with benign programs. A high-priority alarm is raised and a detailed report is sent for inspection.

\bheading{Case 2:} The attack detector recognizes it as an attack, and the benign program detector does not report it as a benign program. This case indicates clear attacks and a high-priority alarm is raised and a detailed report is sent for inspection.

\bheading{Case 3:} The attack detector does not report it as an attack, and the benign program detector recognizes it as a benign program. In this case, the previously detected anomaly is caused by a benign program. The cloud workload is resumed to execute and no alarm is raised.

\bheading{Case 4:} The attack detector does not report it as a known attack, and the benign program detector does not report it as a benign program. In this case, a potential zero-day attack or an unknown benign program is possible. A medium-priority alarm is raised by CloudShield. The cyber analysts can handle these alarms after the high-priority alarms. In fact, in our experiments, we show that case 4 is very unlikely.

\bheading{Response.} Once an anomaly is detected (step 1), CloudShield has already paused the normal cloud workload to shield it from the attacks. Access to highly sensitive data, code, and resources can also be denied, depending on the server's security response policy. If in the second step, an attack is detected, an alarm will be raised. Further responses can be taken to protect the system, and the code and data on it. CloudShield can also stop all processes running on the core. Meanwhile, CloudShield records the relative information into logs for further investigation.

\subsection{System Update} \label{sec:cs:system-update}

We discuss possible system updates of CloudShield. Specifically, CloudShield can update itself if new types of cloud workloads are added, new attacks are discovered or new benign programs are certified. A new model has to be trained only if new cloud workloads are added. For new attacks and benign programs, only the KDE detectors for attacks and benign programs need to be updated.

\bheading{New types of cloud workloads.} The commonly used cloud workloads in practice share common characteristics \cite{khan2012workload, mishra2010towards}, thus this re-training process only needs to be performed when a new type of cloud workload is added. This kind of update is not frequent. Moreover, the whole update procedure can be performed during low usage time. CloudShield loads the updated models and detectors to the processor cores.


\bheading{New certified benign programs.} Update of new certified benign programs is relatively lightweight, compared to cloud workload update, because the pre-trained model does not need to change. CloudShield then executes the new benign program, collects its behavior measurements in a clean execution environment, and calculate the REDs. As shown in the formula of KDE estimator (Eq. \ref{equ:kde}), the estimated likelihood $\hat{f}(x)$ is summed over all reference prediction errors $x_i$. Therefore, CloudShield only needs to append the new prediction errors of the new certified program to the existing prediction errors to form the new RED.


\bheading{New discovered attacks.} This follows the same procedure of updating certified benign programs. It is also lightweight as the pretrained model does not need to be updated.


\section{Evaluation} \label{sec:exp}

\subsection{Experimental Settings} \label{sec:exp:settings}

\bheading{Platform.} We perform our evaluation of CloudShield on a server equipped with 2 Intel Xeon E5-2667 CPUs, each with 6 physical processor cores. Each core has a 32KB L1D (Level-1 Data) cache and a 32KB L1I (Level-1 Instruction) cache. Each package of six cores shares a 256KB L2 (Level-2) cache and a distributed last-level cache of 15MB (2.5MB*6). The server has 64GB memory and a 2TB hard disk. The machine is also equipped with an Nvidia 1080Ti GPU. The HPC values are collected every 10 milliseconds using \textit{Perf} \cite{perf} supplied by the Ubuntu 14.04.6.

\bheading{Cloud workload benchmarks.} We choose five representative cloud benchmarks, as shown in Table \ref{tab:exp:cloudworkloads}. 

\begin{table}[h]
\centering
\caption{Cloud workload benchmarks.} \label{tab:exp:cloudworkloads}
\resizebox{\linewidth}{!}{
\begin{tabular}{|c|l|}
\hline
\textbf{Cloud workload}  & \multicolumn{1}{c|}{\textbf{Description}}          \\ \hline
Web server (Nginx)       & \begin{tabular}[c]{@{}l@{}}Serving 1000 remote connections to request webpages\\ using WRK benchmark \cite{wrk}\end{tabular}                                \\ \hline
Database server (Mysql)  & Performing 128 concurrent queries using SysBench \cite{sysbench}                                                               \\ \hline
Stream server (FFserver) & \begin{tabular}[c]{@{}l@{}}Streaming a MPEG video in real-time to a remote user\\ with FFserver and FFmpeg\end{tabular}                                  \\ \hline
ML training (Pytorch)    & Training an LSTM model using an Nvidia 1080Ti GPU                                                                           \\ \hline
Hadoop                   & Perform Terasort \cite{terasort} using MapReduce                                                                     \\ \hline
\end{tabular}
}
\end{table}

\bheading{Evaluated attacks.} We select nine representative runtime attacks against cloud computing systems for evaluation (Table \ref{tab:exp:evaluated-attacks}). The evaluated attacks are cache side-channel attacks, speculative execution attacks, and buffer overflow attacks. The cache side-channel attacks silently leak information. The four recently discovered speculative execution attacks represent the main hardware resources exploited by the different speculative attack variants. We also evaluate a representative software attack, i.e., buffer overflow attack.



\begin{table}[h]
\centering
\caption{Three catagories of nine attacks are evaluated: cache side-channel attacks, speculative execution attacks and buffer overflow attack.} \label{tab:exp:evaluated-attacks}
\resizebox{0.8\linewidth}{!}{
\begin{tabular}{|c|l|}
\hline
\textbf{Catagory}       & \multicolumn{1}{c|}{\textbf{Attack}}                  \\ \hline
\multirow{4}{*}{\textbf{\begin{tabular}[c]{@{}c@{}}Cache side-\\ channel attacks\end{tabular}}}
        & L1 cache prime-probe attack (l1pp) \cite{gullasch2011cache}           \\ \cline{2-2}
        & L3 cache prime-probe attack (l3pp) \cite{liu2015last}                 \\ \cline{2-2}
        & Flush-reload (fr)                  \cite{yarom2014flush}              \\ \cline{2-2}
        & Flush-flush (ff)                   \cite{gruss2016flush}              \\ \hline
\multirow{4}{*}{\textbf{\begin{tabular}[c]{@{}c@{}}Speculative\\ execution attacks\end{tabular}}}
        & Speculative boundary bypass (spectre v1)   \cite{kocher2019spectre}   \\ \cline{2-2}
        & Indirect branch mis-prediction (spectre v2) \cite{kocher2019spectre}  \\ \cline{2-2}
        & Meltdown (spectre v3)                       \cite{lipp2018meltdown}   \\ \cline{2-2}
        & Speculative store bypass (spectre v4)       \cite{Spectrev4}          \\ \hline
\textbf{Buffer overflow}
        & Stack overflow attack                       \cite{wang2008sigfree}    \\ \hline
\end{tabular}
}
\end{table}


\bheading{Benign programs.} We choose representative benign programs from the SPEC2006 benchmark suite \cite{henning2006spec}. The evaluated benign programs cover a large scope of programs: crypto software (gpg-rsa), compiler (gcc), file and video compression tools (bzip2, h264ref), scientific computation (mcf, milc, namd, libquantum), statistics, and machine learning (soplex, hmmer) and gaming (gobmk).

\bheading{Data collection.} Data were collected in different scenarios. To evaluate the first step, i.e., for detection, we collected data when \ding{172} only the cloud workload is running; \ding{173} the cloud workload is running with benign programs listed above; \ding{174} the cloud workload is running with the attacks listed above; and \ding{175} the cloud workload is running with both benign programs and attacks. To evaluate the second step, i.e., for detection of attacks and benign programs, which we do when the cloud workload is not running, we collected data when \ding{172} only an attack is running; \ding{173} only a benign program is running; and \ding{174} an attack is running together with a benign program. Due to the large number of combinations of cloud workloads, attacks and benign programs, we run each combination for six minutes on a server, and split the data equally into training, validation and testing sets.

\subsection{Metrics}

We first compute an anomaly score for each behavior measurement and then use a threshold to determine False Positive Rate (FPR) and False Negative Rate (FNR).

\bheading{Anomaly score.} An anomaly score is $-log(\hat{f}(x))$, where $\hat{f}(x)$ is the KDE density in Eq. \ref{equ:kde}. Low density $f(x)$ indicates a high anomaly score. During inference, an anomaly score is computed for each behavior measurement, and the score is compared to a threshold to make a binary decision whether it is normal or abnormal (for the normal cloud workload detector), or whether it is a benign program (for the benign program detector), or whether it is an attack (for the attack detector).

\noindent \bheading{Threshold.} The threshold calculation for the cloud workload detector is different from the attack and benign program detectors. The threshold of the cloud workload detector is obtained such that 80\% of the validation normal measurements during the training phase are correctly classified as normal (no attack data are used to construct the normal cloud workload detector and to determine the threshold). For the benign program detector and attack detector, the threshold is obtained such that the FPR (False Positive Rate) and FNR (False Negative Rate) are equal, i.e., the equal error rate is achieved, on the validation set.

\noindent \bheading{False Positive Rate (FPR) and False Negative Rate (FNR).} We also report the standard FPR and FNR as metrics. Based on the anomaly scores and thresholds, the cloud workload detector makes binary decisions (normal or abnormal). Similarly, the benign program detector and known attack detector determine if a benign program or a known attack is running. We report the FPR when the cloud workload or benign programs are running, and the FNR if an attack is running (with and without cloud workloads or benign programs).

\subsection{Anomaly+Attack Evaluation} \label{sec:exp:e2e}

\begin{figure}[h]
    \centering
    \includegraphics[width=\linewidth]{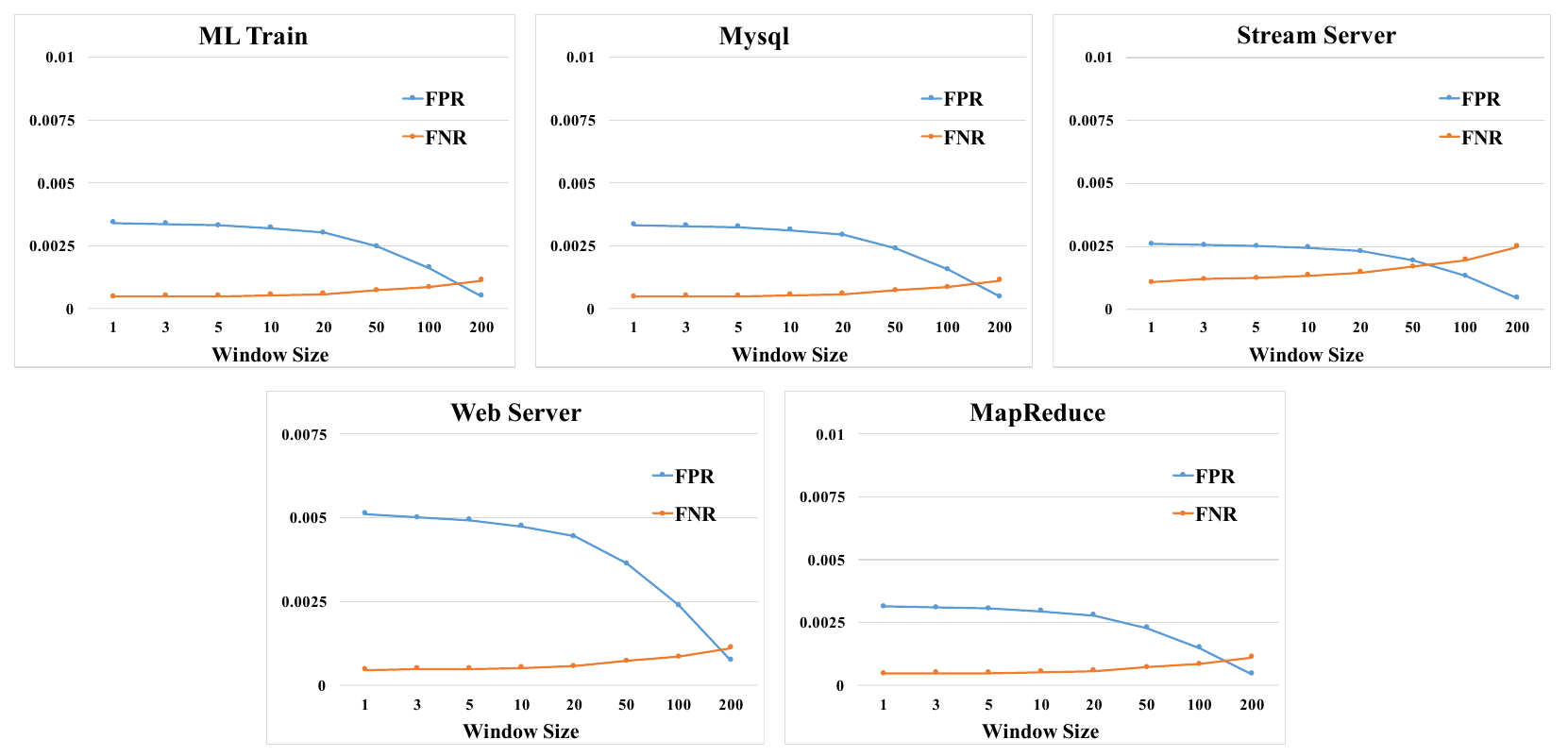}
    \caption{End-to-end (anomaly detection + attack detection) results for 5 different cloud workloads.} \label{fig:exp:window}
\end{figure}

As CloudShield first detects anomalies (step 1) and then identifies attacks and benign programs (step 2), we first illustrate the end-to-end (anomaly detection + attack detection) results in Figure \ref{fig:exp:window} and show the numerical results in Table \ref{fig:exp:window-values}. Separated results and analysis of each step are discussed in Section \ref{sec:exp:realtime}-Section \ref{sec:exp:benign-attack}.

We evaluate different window sizes: if the window size is $w$, in step 1, $w$ contiguous anomalous behavior marker measurements are identified as an anomaly. Similarly, $w$ contiguous behavior marker measurements are collected before an attack or benign program can be identified. For a specific cloud workload, we report the average FPR for that cloud workload + each benign program. We report the average FNR for that cloud workload + each attack + each benign program we evaluated. A higher FPR increases the number of false alarms, while a higher FNR increases the chance that an attack will go undetected. Low rates of both are desired. We observe that the CloudShield indeed has very low FNRs for all 5 workloads for all window sizes - less than 0.3\%, indicating excellent detection accuracy and hence, excellent security. FPRs are slightly higher but also less than 0.6\%. When $w$=1, the webserver workload has the highest FPR (0.51\%), while stream server achieves the lowest FPR (0.26\%). For all five cloud workloads, the FPR decreases as $w$ becomes larger, however, the FNR increases accordingly. When $w=100$, FPR decreases to 0.13\% (for stream server) and 0.24\% (for webserver). FNR increases to 0.09\% and 0.19\%. When $w=200$, FNR tends to exceed FPR for all five cloud workloads. Note that a larger window size can increase the detection delays (evaluated in Section \ref{sec:exp:performance}). Because of the low FPR and FNR, a window size of 5-10 should be sufficient.

\begin{table}[h]
    \centering
    \caption{Quantitative end-to-end (anomaly detection + attack detection) evaluation results.} \label{fig:exp:window-values}
    \includegraphics[width=0.8\linewidth]{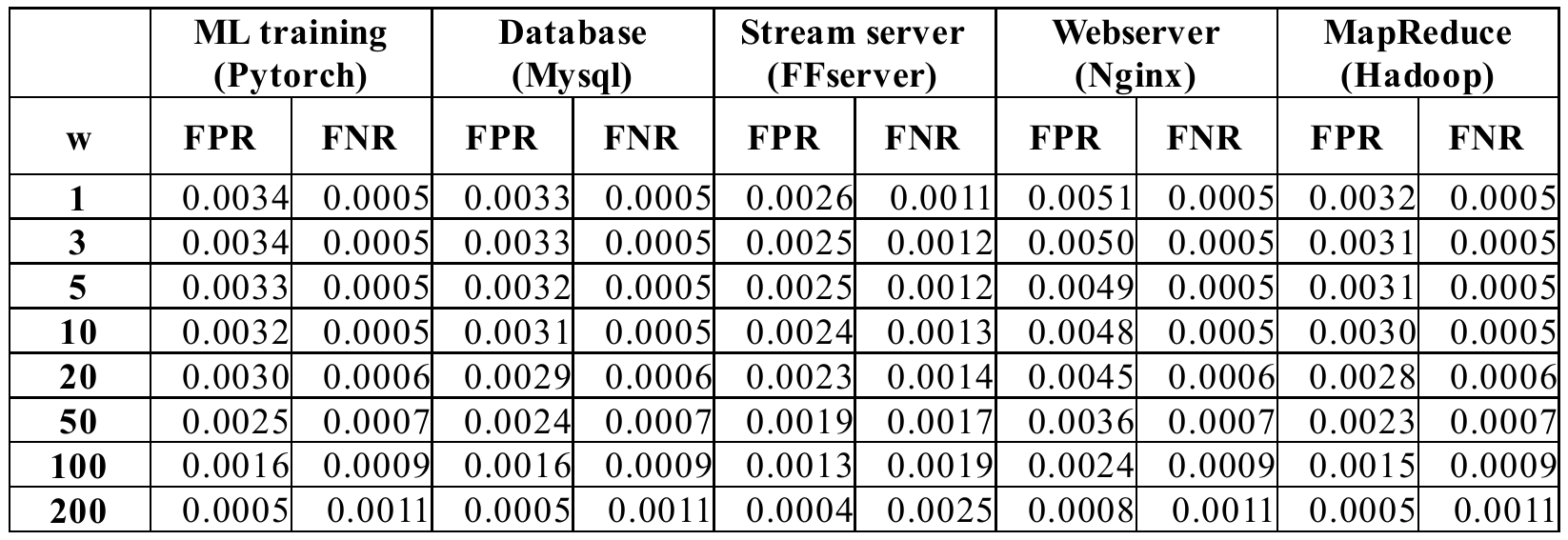}
\end{table}

We compare the proposed CloudShield to four representative anomaly detection methods in the literature, i.e., Isolation Forrest (IF) \cite{liu2008isolation}, One-class SVM (OCSVM) \cite{scholkopf2000support}, Local Outlier Factor (LOF) \cite{breunig2000lof}, and Principal Component Analysis (PCA) \cite{hotelling1933analysis}. We show the end-to-end results in Table \ref{tab:exp:previous-comparison}. For the existing anomaly detection methods, we replace the pretrained model + KDE in steps 1 and 2 of CloudShield with the corresponding method. We average the FPR and FNR across each combination of cloud workload, benign program, and attack. We observe that, with $w$=5 or $w$=10, CloudShield achieves lower FPR and FNR compared to other methods. Specifically, when $w$=5, the best FNR and FPR of existing methods are 1.41\% (OCSVM) and 6.95\% (PCA), respectively, while CloudShiled has much lower (better) FPR of 0.34\% and FNR of 0.06\%. Similar results are shown when $w$=10.

{\textcolor{red}{
\begin{table}[h]
\centering
\caption{Compare CloudShield to existing anomaly detection methods.}\label{tab:exp:previous-comparison}
\resizebox{\linewidth}{!}{
\begin{tabular}{|c|l|c|c|}
\hline
                               &                                     & \textbf{False Positive Rate (FPR)} & \textbf{False Negative Rate (FNR)} \\ \hline
\multirow{5}{*}{\textbf{w=5}}  & \textbf{Isolation Forrest (IF)}     & 0.1728                             & 0.442                              \\ \cline{2-4}
                               & \textbf{One-class SVM (OCSVM)}      & 0.0141                             & 0.1011                             \\ \cline{2-4}
                               & \textbf{Local Outlier Factor (LOF)} & 0.0518                             & 0.0956                             \\ \cline{2-4}
                               & \textbf{PCA}                        & 0.0587                             & 0.0695                             \\ \cline{2-4}
                               & \textbf{CloudShield}                & 0.0034                             & 0.0006                             \\ \hline
\multirow{5}{*}{\textbf{w=10}} & \textbf{Isolation Forrest (IF)}     & 0.1539                             & 0.416                              \\ \cline{2-4}
                               & \textbf{One-class SVM (OCSVM)}      & 0.01571                            & 0.1031                             \\ \cline{2-4}
                               & \textbf{Local Outlier Factor (LOF)} & 0.0516                             & 0.0990                             \\ \cline{2-4}
                               & \textbf{PCA}                        & 0.0519                             & 0.1150                             \\ \cline{2-4}
                               & \textbf{CloudShield}                & 0.0033                             & 0.0007                             \\ \hline
\end{tabular}
}
\end{table}
}}

\subsection{Can CloudShield Detect Anomalous Behavior in Realtime?} \label{sec:exp:realtime}

A key challenge for real-time anomaly detection is short or stealthy attacks. Attacks can hide by switching between running and sleeping. A good anomaly detection system should be able to capture the attack once it is running. We evaluate CloudShield against such attacks and show it can detect them almost immediately. We schedule each of the nine attacks to run and then sleep for a random period (10s-40s) before the next attack runs. The experiment is performed when the ML training workload is running.

\begin{figure}[h]
    \centering
    \includegraphics[width=0.7\linewidth]{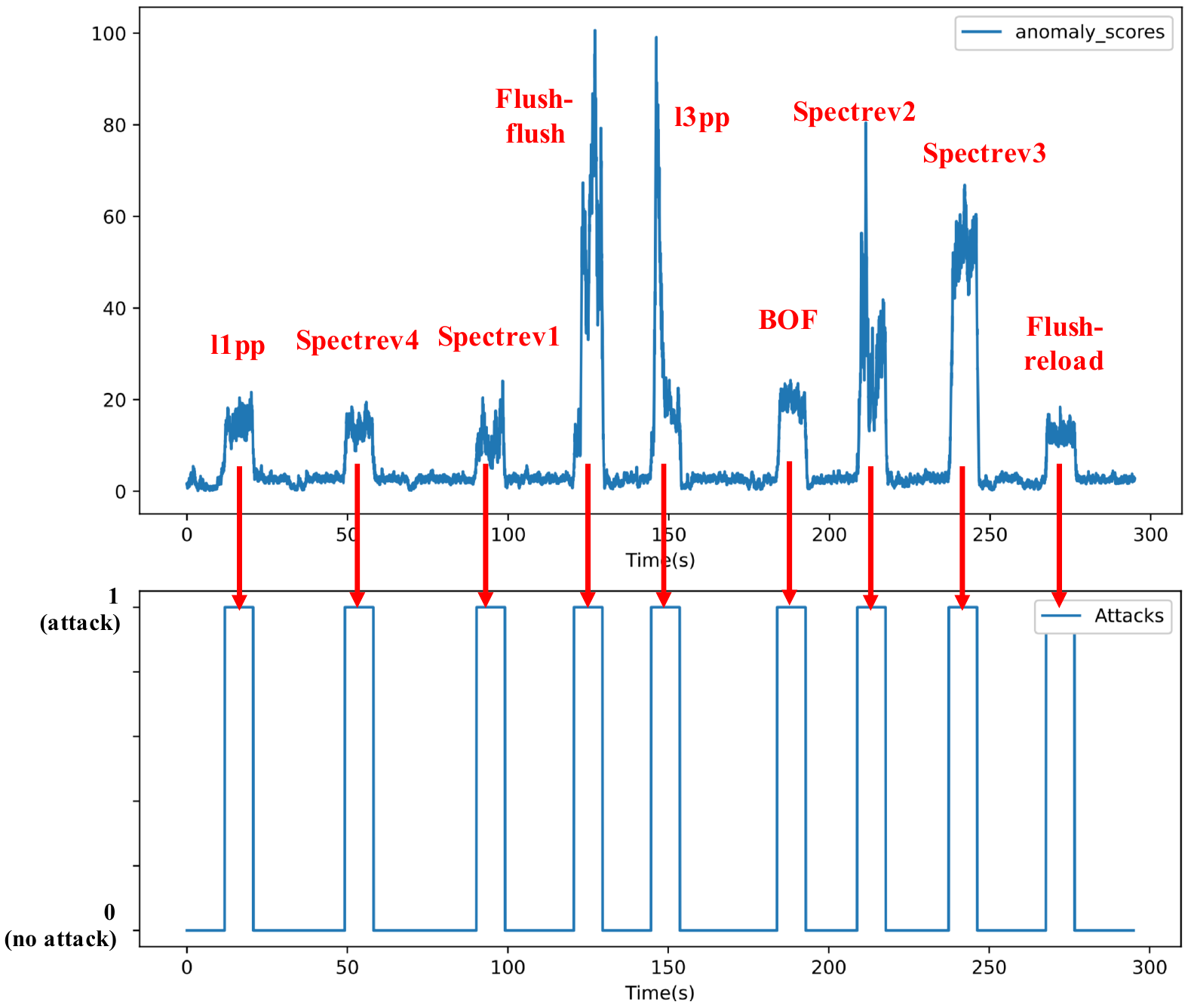}
    \caption{Real-time detection of anomalies. We schedule each of the nine attacks to run for 10 seconds and sleep for a random period of time (10-40s).} \label{fig:detection}
\end{figure}

We show the attack scheduling and the anomaly scores output ($-log(\hat{f}(x))$ in Eq. \ref{equ:kde}) by CloudShield in Figure \ref{fig:detection}. It is clear that once an attack is running, possibly after sleeping, CloudShield captures it (indicated by a large anomaly score). Once the attack program's behavior is suspended, the anomaly score quickly goes back to a low value. Therefore, the proposed CloudShield can detect anomalies in real-time. We also observe that two last-level cache attacks, i.e., the flush-flush attack (ff) and the LLC prime-probe attack (l3pp), and two speculative execution attack variants, i.e., the spectre v2 and the spectre v3, result in higher anomaly scores than the other attacks, indicating their distinctive behavior.

\subsection{Can CloudShield Detect Zero-day Attacks?} \label{sec:exp:zero-day}

We evaluate the anomaly detection (step 1) on the nine attacks, including the four recently proposed speculative execution attacks. \textit{Note that in the anomaly detection step, CloudShield is only trained on the normal behavior of the cloud workloads, and has not seen code or data of any of the nine attacks, so they are like zero-day attacks to CloudShield in this experiment.}

We consider the model predictions for the four scenarios:
\begin{enumerate}
    \item Normal workload
    \item Normal workload and a benign program running
    \item Normal workload and an attack running
    \item Normal workload, a victim program, and an attack running
\end{enumerate}

\begin{figure}[h]
    \centering
    \includegraphics[width=0.7\linewidth]{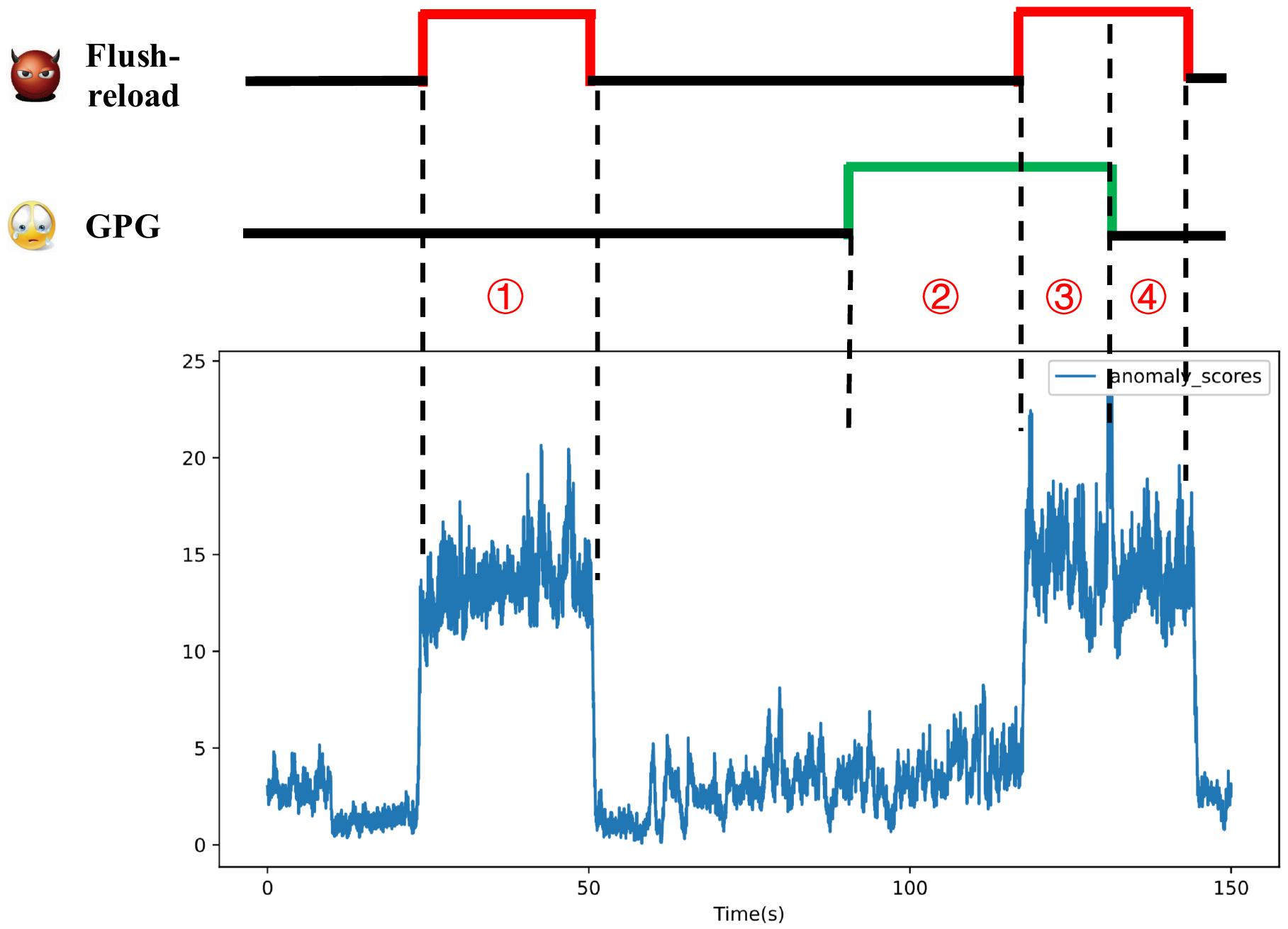}
    \caption{An example of anomaly detection of the flush-reload attack with the ML training workload.} \label{fig:exp:detection-gpg-fr}
\end{figure}

We first illustrate a real example of anomaly detection in Figure \ref{fig:exp:detection-gpg-fr} with the ML training workload. We run a flush-reload attack and a victim program, i.e., gpg-rsa. In period \ding{172}, the flush-reload attack is activated. We observe that the anomaly score quickly increases significantly. In period \ding{173}, the victim program gpg-rsa is running and it was not recognized as an anomaly. In the period \ding{174}, the flush-reload attack is executed and the anomaly score again quickly jumps to a high value. In period \ding{175}, the victim program ends and the anomaly score remains high as the attack is still running.

We show quantitative results of anomaly detection (step 1) in Table \ref{tab:exp:ad}. The first line of the results is scenario 1 where only the normal cloud workload is running. The next four lines show scenario 2, i.e., the normal workload and an additional benign program are running. The next nine lines present the results of scenario 3, where an attack is running concurrently with the normal cloud workload. Finally, representatives of scenario 4 are shown in the remaining (27) lines of Table \ref{tab:exp:ad} where an attack and a victim program are running together.


\begin{table}[h]
    \centering
    \caption{Results of anomaly detection (step 1) with different cloud workloads.}\label{tab:exp:ad}
    \includegraphics[width=\linewidth]{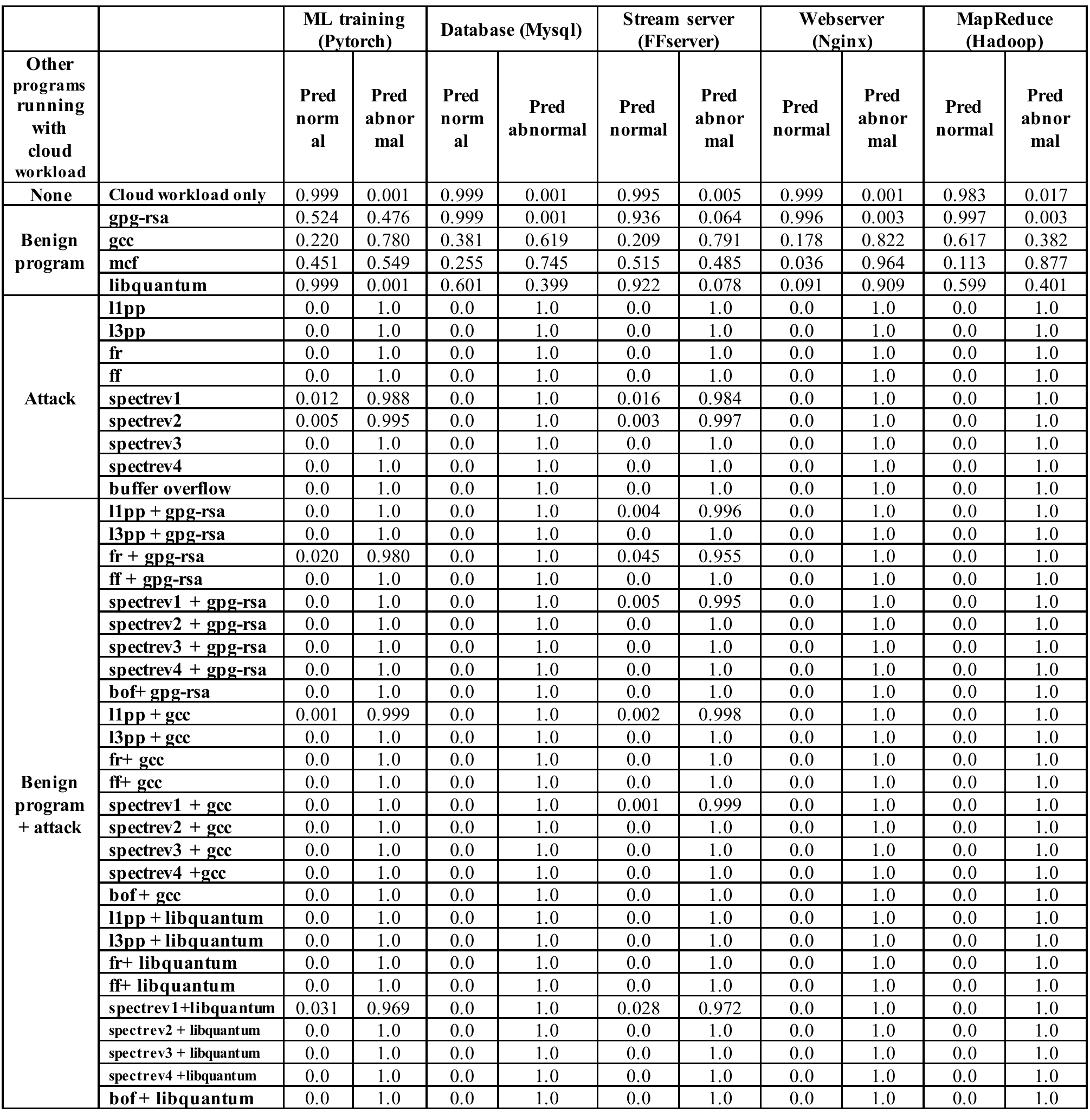}
\end{table}

We find that when only the normal workload is running (scenario 1), CloudShield almost always correctly recognizes it as normal (the first line) for the ML training, database, stream server, and web server benchmarks with a 0.1\%-0.5\% false positive rate (predict abnormal column). When MapReduce is running, CloudShield misrecognizes 1.7\% of normal workloads as anomalous -- still a small level.

When a benign program is running concurrently with the cloud benchmark (scenario 2), we observe that the results highly depend on the cloud workload and the benign program. For example, the GPG-RSA is recognized as normal with less than 1\% false positive rate in the database, web server, and MapReduce workloads. However, large false positives, i.e., 6.4\% and 47.6\% of the GPG-RSA, are observed in the stream server and ML training workloads, respectively. These false alarms can cause false alarm fatigue. Thus it requires the next step to further distinguish benign programs versus malicious anomalies, and reduce the number of false alarms. Note that CloudShield distinguishes certified benign programs from attacks (step 2) to reduce false alarms after an anomaly is identified (results discussed in Section \ref{sec:exp:benign-attack}).

In scenario 3, once an attack is running with the cloud workload, it can be detected with zero false negatives in the database, web server, and MapReduce workloads. For the ML training workload, the Spectre v1 and v2 attacks cause 1.2\% and 0.5\% false-negative rates, respectively. For the stream server workload, the Spectre v1 and v2 attacks introduce 1.6\% and 0.3\% false-negative rates, respectively. These results show that CloudShield is capable of detecting zero-day attacks, since the normal cloud workload detector in step1 has not been trained with any attack.

We also evaluate scenario 4 where an attack program runs concurrently with a benign or victim program and the cloud workload. Similar to scenario 3, we observe that the attacks can be detected with zero false negatives with the database, web server, and MapReduce workloads. For the ML training workload, the worst case is when spectre v1 and libquantum are running concurrently, the attack is missed by 3.1\%, slightly higher than scenario 3 where the spectre v1 attack is running alone (1.2\%). For the stream server workload, the highest false-negative rate is 4.5\% when a flush-reload attack is executed with gpg-rsa. Next is when Spectre v1 and libquantum are running with the stream server, the FNR is 2.8\%. Although these results from just step 1 for anomaly detection are very good for not missing attacks (low FNRs), the FPRs in scenario 2 when benign programs cause false alarms seem higher than we would like to see. Hence, we propose step 2, to detect benign anomalies from real attacks.

\subsection{Can CloudShield Distinguish Benign Anomalies from Attacks?} \label{sec:exp:benign-attack}

Anomalies can be caused by benign programs, i.e., benign anomalies. Therefore, once an anomaly is detected, CloudShield takes the next step to figure out whether it is a benign anomaly or an attack. As shown earlier, CloudShield implements two detectors to identify known attacks and certified benign programs, respectively. These two detectors can reduce false alarms by 99.0\%.

\begin{figure}[h]
    \centering
    \includegraphics[width=\linewidth]{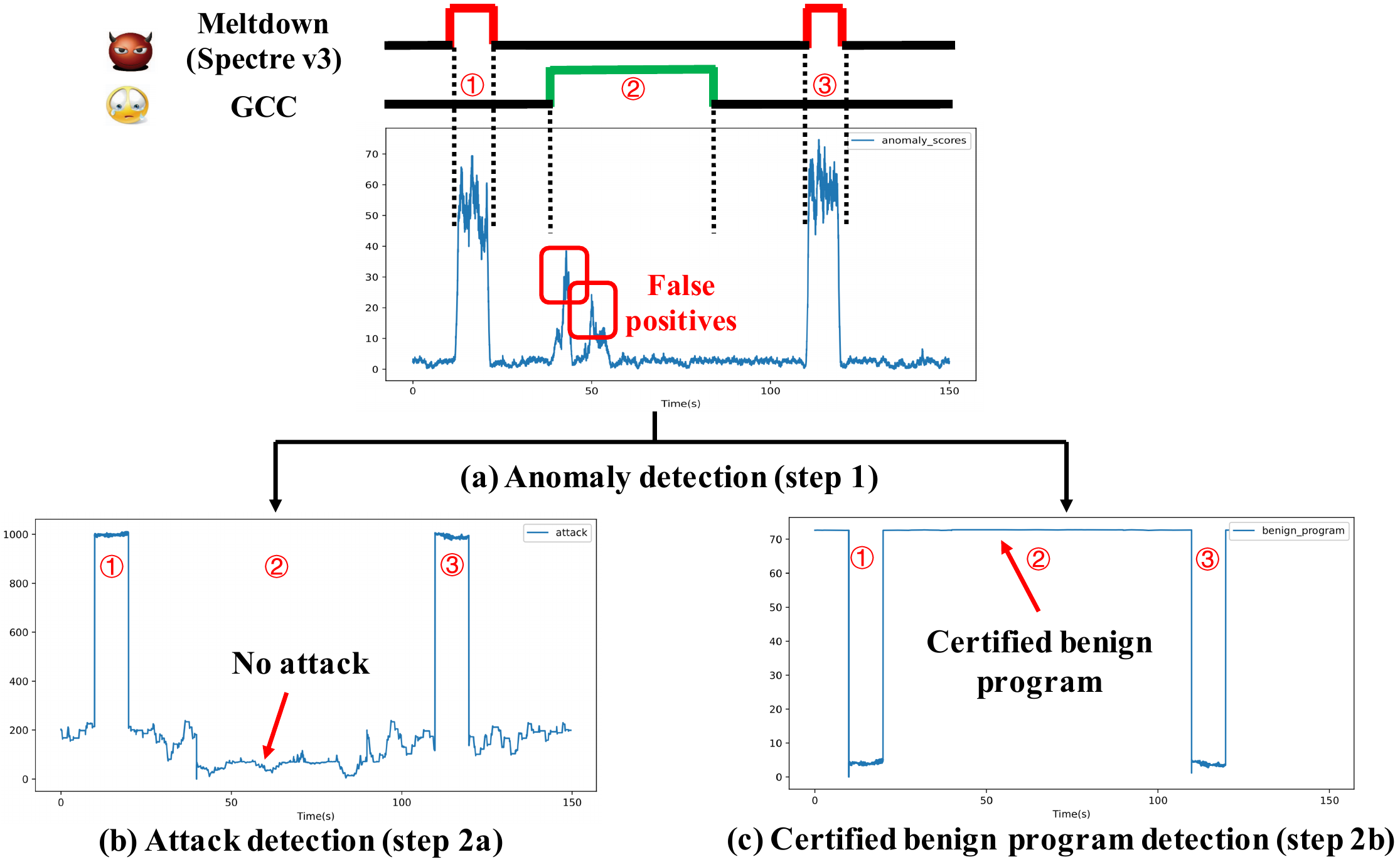}
    \caption{An example of reducing false alarms by indentifying attacks and certified benign programs.} \label{fig:exp:gcc-spectrev3}
\end{figure}

We show a real example of CloudShield reducing false alarms by distinguishing known attacks and certified benign programs in Figure \ref{fig:exp:gcc-spectrev3}. We run an attack (spectre v3) and a benign program (gcc), both with the ML training workload. The periods \ding{172} and \ding{174} indicate that the attack is running, and the period \ding{173} means the benign program is running. Figure \ref{fig:exp:gcc-spectrev3} (a) illustrates the anomaly scores in the anomaly detection step. We observe that while both attacks are correctly identified (periods \ding{172} and \ding{174}), the beginning of gcc execution is incorrectly recognized as attacks (false alarms). Then the ML training workload is paused and the behavior measurements are re-collected as input to the two step 2 detectors. Figure \ref{fig:exp:gcc-spectrev3} (b) shows the result of the attack detector. High values indicate an attack and low values mean no attack. It correctly identifies periods \ding{172} and \ding{174} as attacks, while \ding{173} is not an attack. Figure \ref{fig:exp:gcc-spectrev3} (c) shows the result of the benign program detector. High values represent a benign program and low values indicate a program that is not in the set of certified benign programs. We find that the certified benign program detector reports high values in period \ding{173} (and idle periods), while the values in periods \ding{172} and \ding{174} are low (not certified benign programs). Jointly considering the two detectors, CloudShield correctly determines that \ding{173} is a certified benign program, while \ding{172} and \ding{174} are real attacks.



We show quantitative results of attacks and certified benign program detection (step 2) in Table \ref{tab:exp:benign-attack-w5}. We select eleven representative benign programs from the SPEC benchmark suite and the same nine attacks as in previous sections for evaluation. For the benign program detection, we observe that six benign programs (gpg-rsa, bzip2, namd, soplex, hmmer, and libquantum) can be recognized correctly with no false alarms. The milc program introduces the highest but acceptable FPR of 3.5\%. Of this, 2.6\% were identified as stealthy attacks and 0.9\% as zero-day attacks or unknown benign programs. On average, 99.0\% of the benign programs can be identified correctly, i.e., the false alarms raised by benign programs in the anomaly detection is suppressed by 99.0\%. Within the remaining false alarms (1.0\%), we observe that 0.6\% are recognized as case 4 (zero-day attacks or unknown benign programs) which results in a medium-priority alarm, and 0.4\% are recognized as high-priority attacks (case 1 and 2). For attack detection, we observe that all attacks are correctly identified. A detailed analysis shows that 99.8\% of attacks are identified as high-priority attacks (case 2) and 0.2\% attacks are recognized as stealthy attacks.


\begin{table}[h]
    \centering
    \caption{Results of benign programs/attacks detection (step 2).} \label{tab:exp:benign-attack-w5}
    \includegraphics[width=\linewidth]{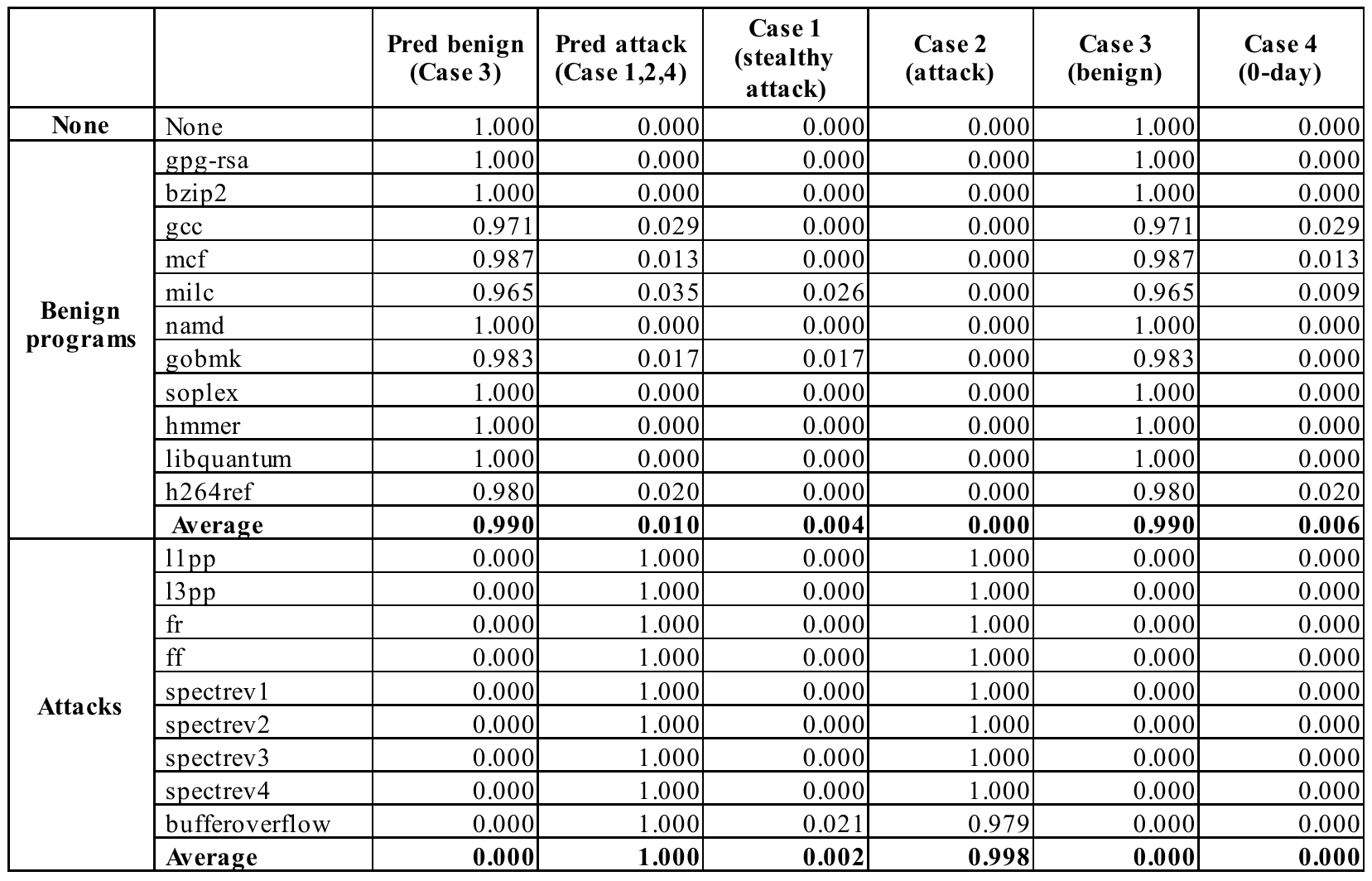}
\end{table}

\begin{table}[h]
    \centering
    \caption{Results of benign programs/attacks detection (step 2), when attacks and benign programs run concurrently.} \label{tab:exp:benign-attack-concurrent-w5}
    \includegraphics[width=\linewidth]{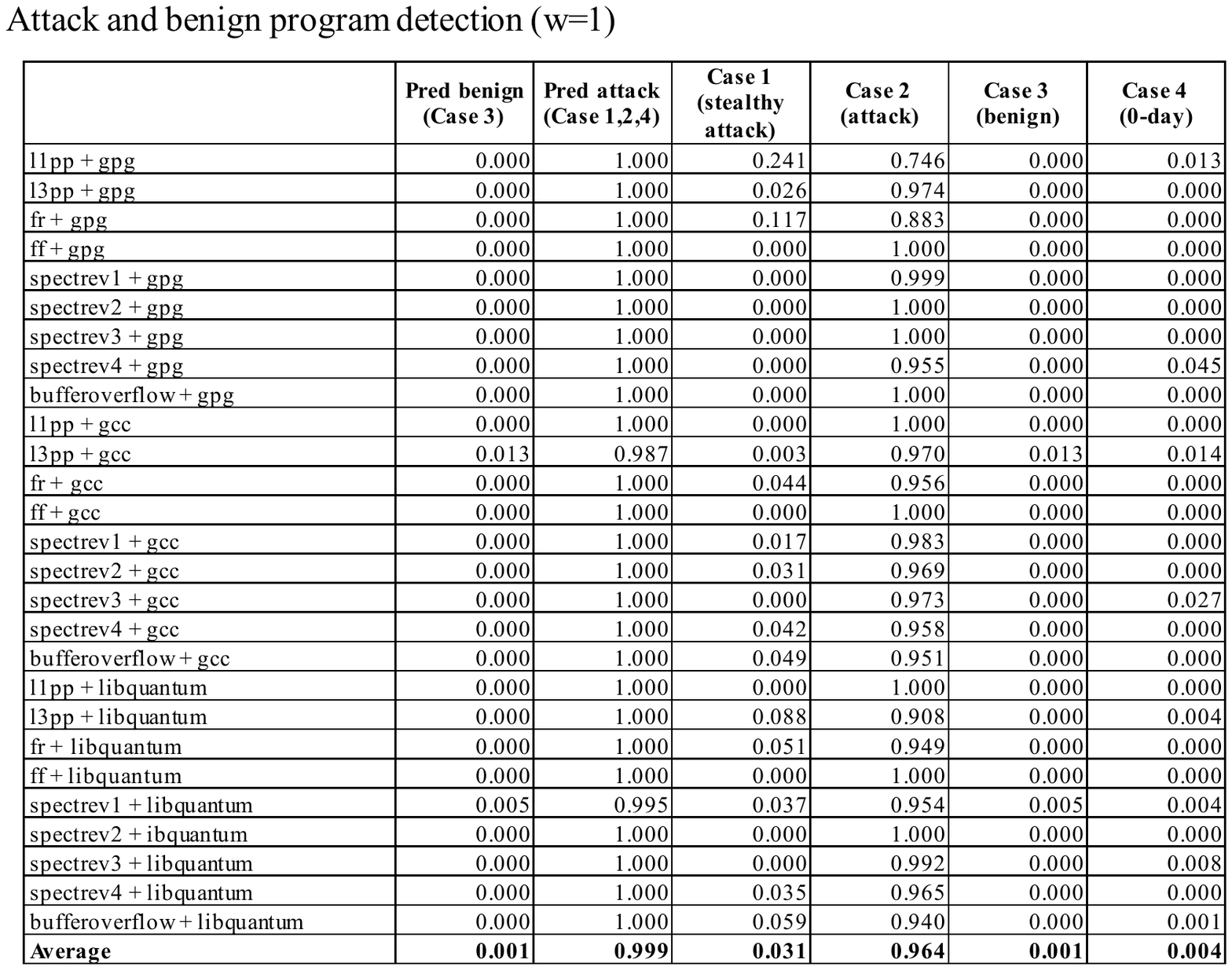}
\end{table}

We consider a more difficult scenario where an attack is running concurrently with a certified benign program. We run three benign programs: gpg-rsa, gcc, and libquantum with the nine evaluated attacks in Table \ref{tab:exp:benign-attack-concurrent-w5}. First, on average, 99.9\% of attacks when they are concurrently running with benign programs, are correctly recognized as attacks. A detailed analysis shows that, when an attack program is running concurrently with a benign program, 96.4\% are identified as high-priority attacks (case 2), 3.1\% are recognized as high-priority stealthy attacks (case 1), only 0.4\% are classified as medium-priority zero-day attacks (case 4). These results show that CloudShield can still detect attacks even if they hide in benign programs.

\begin{table}[h]
    \centering
    \caption{Results of zero-day (unknown) attacks in step 2.} \label{tab:exp:zero-day-attack}
    \includegraphics[width=\linewidth]{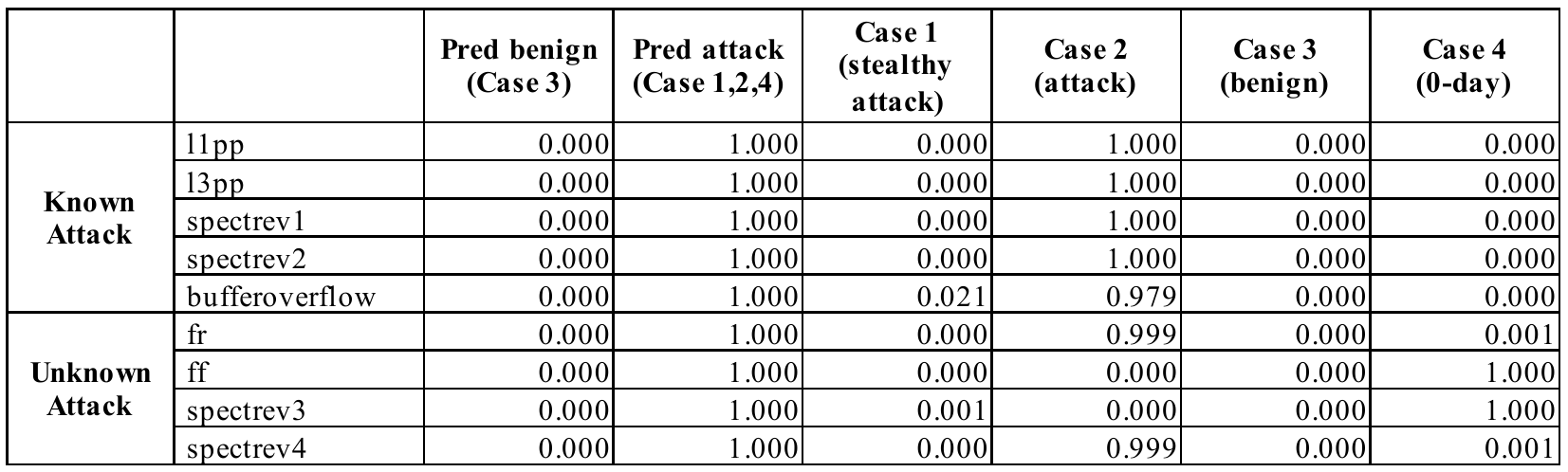}
\end{table}

\bheading{Zero-day attack detection in step 2.} We conduct another experiment by putting only L1 prime-probe (l1pp), LLC prime-probe (l3pp), spectre v1, spectre v2, and buffer overflow attacks in the set of known attacks. This means that the flush-reload (fr), flush-flush (ff), spectre v3, and spectre v4 attacks are unknown zero-day attacks. We show the known and zero-day attack detection results in Table \ref{tab:exp:zero-day-attack}. We observe that CloudShield can still correctly recognize unknown attacks. The flush-flush and spectre v3 attacks are classified as case 4 (zero-day attacks). The other two attacks, i.e., the flush-reload and spectre v4 attacks, are detected as known attacks probably because their behavior is similar to the known attacks. 

\bheading{Necessity of the two-step method.} We have also investigated detecting attacks together with detecting anomalies in the first step, when the cloud workloads are running. The benefit of doing this is that the attacks can be identified more quickly, than in the second step. However, the downside of detecting attacks in the first step is that the attacks and cloud workloads interfere with each other, making the behavior markers collected in the first step not capable enough to identify the attacks. Hence our two-steps method is much better.

\subsection{Detection Latency and Overhead} \label{sec:exp:performance}

\begin{figure}[h]
    \centering
    \includegraphics[width=0.8\linewidth]{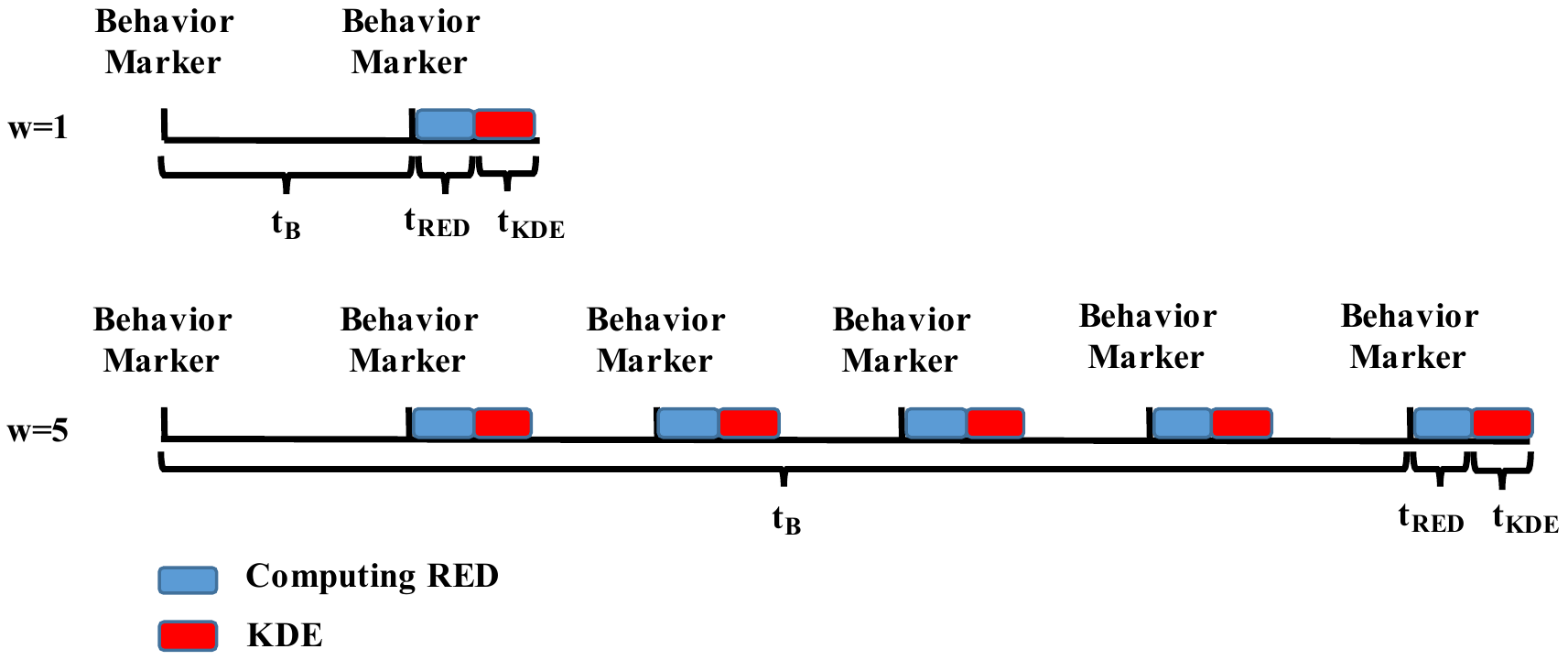}
    \caption{Illustration of the timeline for anomaly detection.} \label{fig:exp:timeline}
\end{figure}

\bheading{Detection latency.} The detection latency is defined as the period from the time the attack starts running, to the time an attack alarm is raised. We present the overhead of robust detection using more than one set of behavior marker measurements, e.g., with a sequence of $w=5$ sets of measurements. The timeline for detecting an attack is shown in Figure \ref{fig:exp:timeline} (similar for attack and benign program detection). $t_{B}$ denotes the time interval for collecting $w$ behavior marker measurement. $t_{RED}$ represents the time needed for computing the RED by inferencing the pre-trained model. $t_{KDE}$ is the time to infer the KDE detector. The computation of RED and KDE can overlap with the behavior marker collection if $w>1$ (Figure \ref{fig:exp:timeline}).

Table \ref{tab:exp:latency} presents the detection latency when $w=$1, 5, 10, 50 and 100. As the HPCs are sampled every 10ms, $t_B$=10ms when $w$=1. We measure $t_{RED}$ and $t_{KDE}$ on the server. Specifically, the calculation of RED ($t_{RED}$) is performed on the GPU and the calculation of KDE ($t_{KDE}$) is performed on the CPU of the server. The two overall numbers in the parenthesis are detection time when there is no anomaly (thus no step 2) and there is an attack, respectively. We show that CloudShield can detect anomalies and identify the attacks and benign programs in 32 to 112 milliseconds if $w=1$ or $w=5$. Considering the attack usually takes seconds to succeed, e.g., several encryption operations for side-channel attacks, this latency can achieve our design goal of \textit{real-time} detection. Larger window sizes can reduce the false-positive rate, however, the false-negative rate is slightly increased (Figure \ref{fig:exp:window}) and detection time is increased to seconds. We suggest $w$=5 is sufficient.

\begin{table}[h]
\centering
\caption{Detection latency (ms) versus window sizes.}\label{tab:exp:latency}
\resizebox{\linewidth}{!}{
\begin{tabular}{|c|c|c|c|c|c|c|c|}
\hline
\textbf{(ms)} & \multicolumn{3}{c|}{\textbf{Anomaly Detection}} & \multicolumn{3}{c|}{\textbf{Benign program/Attack detection}} & \textbf{Overall (no anomaly, attack)} \\ \hline
              & \textbf{$t_B$}   & \textbf{$t_{RED}$}  & \textbf{$t_{KDE}$}  & \textbf{$t_B$}        & \textbf{$t_{RED}$}       & \textbf{$t_{KDE}$}       & \textbf{}        \\ \hline
\textbf{w=1}  & 10.0          & 0.02           & 0.76           & 10.0               & 0.02                & 1.58               & (10.78, 32.38)                                                \\ \hline
\textbf{w=5}  & 50.0          & 0.02           & 0.76           & 50.0               & 0.02                & 1.58               & (50.78, 112.38)                                               \\ \hline
\textbf{w=10}  & 100.0          & 0.02           & 0.77           & 100.0               & 0.02                & 1.60               & (100.79, 212.41)                          \\ \hline
\textbf{w=50}  & 500.0          & 0.02           & 0.78           & 500.0               & 0.02                & 1.62               & (500.80, 1012.44)                         \\ \hline
\textbf{w=100}  & 1000.0          & 0.02           & 0.79           & 1000.0               & 0.02                & 1.65               & (1000.81, 2012.48)                        \\ \hline
\end{tabular}
}
\end{table}

\begin{figure}[h]
    \centering
    \includegraphics[width=0.6\linewidth]{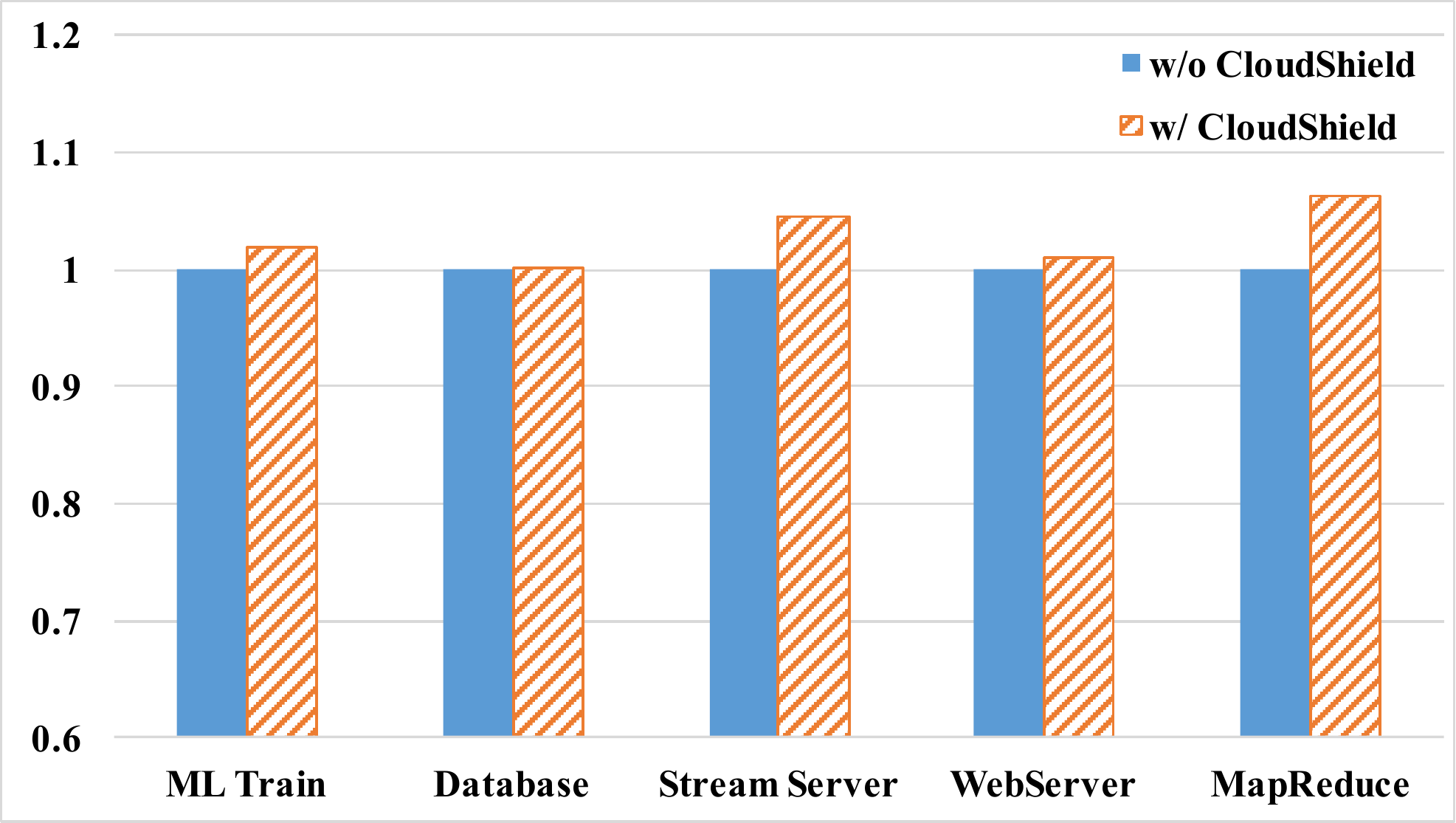}
    \caption{Performance overhead of CloudShield with different cloud workloads.} \label{fig:exp:overhead}
\end{figure}

\bheading{Performance overhead.} We evaluate the performance overhead of CloudShield. We use the benchmarks in Table \ref{tab:exp:cloudworkloads}. We use completion time as the metric for ML training and MapReduce, average time per query for Database and Webserver, and processing time per frame for Stream Server. All the metrics are normalized to the cloud workload running without Cloudshield. Figure \ref{fig:exp:overhead} reports the normalized metrics without CloudShield (blue solid) and with CloudShield running (orange dashed). Results are averaged over five runs). We see that CloudShield only introduces a small performance overhead. The maximum overhead is 6.3\% for MapReduce and the minimum is 0.5\% for database. In our experiments, we observe that on average CloudShield consumes 17.1\% CPU time on the server.


\subsection{Discussion: Evasion Attacks}

There have been many attacks against deep learning systems \cite{he2019sensitive, he2020attacking, he2019model, yao2020miss}. Previous work \cite{qiu2020adversarial} revealed that attackers can effectively generate adversarial examples in the black-box setting to evade deep learning based intrusion detection systems. However, generating adversarial examples against our system is generally harder. First, our system monitors the \textbf{dynamic} behavior of a program. Generating dynamic adversarial examples that can both interact with other programs and escape detection in the black-box setting remains challenging. Second, the behavior markers monitored in our system are HPC measurements. As HPC measurements highly depend on the context of the executing environment, this introduces an extra obstacle for the attacker to construct the same execution environment when generating evasion adversarial examples. How to design and develop efficient evasive attacks and how to detect these attacks are worth exploring as future work.

\section{Past Work} \label{sec:pastwork}

Past work on anomaly detection in the cloud mainly focused on machine performance degradation. Liu \etal \cite{liu2016anomaly} proposed self-organizing maps for detecting anomalies in machine performance. Vallis \etal \cite{vallis2014novel} leveraged statistical measurement, e.g., median, and median absolute deviation, to detect machine performance degradation. Pannu \etal \cite{pannu2012aad} implemented adaptive anomaly detection (AAD) based on non-linear transformation for detecting failures in cloud infrastructures. These work detected performance anomalies to provide reliable cloud service, however, unlike our work, they did not consider anomalies caused by attacks.

Another line of research detected specific attacks in the cloud. For example, Zhang \etal \cite{zhang2016cloudradar} developed CloudRadar for side-channel attack detection in the cloud using hardware performance counters. Guo \etal \cite{guo2020specusym} detected cache side-channel leakage with symbolic execution. Wang \etal \cite{wang2020kleespectre} leveraged symbolic execution to detect speculative execution attacks. However, each of these detected a specific type of attack, unlike our work, which covers a broader scope of attacks, including zero-day attacks.

Recent work used deep learning for anomaly detection. Alam \etal \cite{alam2019zero} proposed AutoPerf based on an autoencoder to detect hardware performance anomalies. DeepLog \cite{du2017deeplog} leveraged system event logs to detect system failures. Sucheta \etal \cite{chauhan2015anomaly} and Malhotra \etal \cite{malhotra2015long} proposed LSTM for sequential anomaly detection. He \etal \cite{he2019power} leveraged LSTM for anomaly detection in critical infrastructures. Du \etal \cite{du2019lifelong} updated anomaly detection model through unlearning. Hu \etal \cite{hu2021smartphone} developed a deep learning hardware module for impostor detector in smartphone systems. As stated in their work, unlearning may introduce a higher false-positive rate. In contrast, CloudShield significantly reduces false positives by distinguishing benign and malicious anomalies.

\ifisthesis
\section{Chapter Summary} \label{sec:conclusion}
\else
\section{Conclusion} \label{sec:conclusion}
\fi

\ifisthesis
In this chapter,
\else
In this paper,
\fi
we proposed CloudShield, a real-time anomaly and attack detection system for cloud computing. CloudShield leverages a single pre-trained deep learning model and
\ifisthesis
extends the reconstruction error distribution (RED) of hardware performance counters in Chapter \ref{ch:power-grid}
\else
leverages the reconstruction error distribution (RED) of hardware performance counters
\fi
to model the normal behavior of a system using kernel density estimation (KDE). It is worth noting that CloudShield explicitly takes false-alarm reduction into account, a critical problem in anomaly detection systems. Once an anomaly is detected, CloudShield automatically distinguishes benign programs, known attacks, and zero-day attacks by investigating the different attack and benign program reconstruction error distributions, using the pre-trained model and kernel density estimators.

We evaluate CloudShield on various cloud workloads, attacks, and benign programs. Experimental results show that CloudShield can reliably detect various attacks in real-time with high accuracy and very low FNR and FPR. Moreover, experiments show that it can correctly identify unknown zero-day attacks and stealthy attacks that are running concurrently with benign programs. CloudShield achieves very low 0.3\% FNR and 0.6\% FPR for overall anomaly-attack detection. Especially, we find that CloudShield can detect the recently proposed speculative execution attacks in 32-112ms, and it can reduce false alarms by up to 99.0\%.

\ifisthesis
In Chapter \ref{ch:power-grid} and Chapter \ref{ch:side-channel} (this chapter), we demonstrate that deep learning is very powerful in solving a critical security problem, i.e., anomaly detection, in the power-grid controller and in cloud computing systems, respectively. However, deep learning models and data have become the targets of attacks. In the following chapters, we provide a taxonomy of deep learning attacks and defenses in Chapter \ref{ch:dl-attack-pastwork}, and we investigate attacks and defenses of deep learning itself. Specifically, we discuss protecting deep learning model integrity in Chapter \ref{ch:integrity} and inference data privacy in Chapter \ref{ch:confidentiality}.
\fi

\bibliographystyle{IEEEtranS}
\bibliography{main}

\begin{thebibliography}{10}
\providecommand{\url}[1]{#1}
\csname url@samestyle\endcsname
\providecommand{\newblock}{\relax}
\providecommand{\bibinfo}[2]{#2}
\providecommand{\BIBentrySTDinterwordspacing}{\spaceskip=0pt\relax}
\providecommand{\BIBentryALTinterwordstretchfactor}{4}
\providecommand{\BIBentryALTinterwordspacing}{\spaceskip=\fontdimen2\font plus
\BIBentryALTinterwordstretchfactor\fontdimen3\font minus
  \fontdimen4\font\relax}
\providecommand{\BIBforeignlanguage}[2]{{%
\expandafter\ifx\csname l@#1\endcsname\relax
\typeout{** WARNING: IEEEtranS.bst: No hyphenation pattern has been}%
\typeout{** loaded for the language `#1'. Using the pattern for}%
\typeout{** the default language instead.}%
\else
\language=\csname l@#1\endcsname
\fi
#2}}
\providecommand{\BIBdecl}{\relax}
\BIBdecl

\bibitem{amazonaws}
\url{https://aws.amazon.com}, 2018.

\bibitem{googlecloud}
\url{http://cloud.google.com}, 2018.

\bibitem{microsoftazure}
\url{https://azure.microsoft.com/en-us/services/machine-learning-studio/},
  2018.

\bibitem{meltdown3a}
\url{https://cve.mitre.org/cgi-bin/cvename.cgi?name=CVE-2018-3640}, 2018.

\bibitem{Spectrev4}
\url{https://msrc-blog.microsoft.com/2018/05/21/analysis-and-mitigation-of-speculative-store-bypass-cve-2018-3639/},
  2018.

\bibitem{wrk}
\url{https://github.com/wg/wrk}, 2019.

\bibitem{sysbench}
\url{https://github.com/akopytov/sysbench}, 2019.

\bibitem{perf}
\url{https://perf.wiki.kernel.org/index.php/Main_Page}, 2020.

\bibitem{terasort}
\url{https://hadoop.apache.org/docs/current/api/org/apache/hadoop/examples/terasort/package-summary.html},
  2020.

\bibitem{alam2019zero}
M.~Alam, J.~Gottschlich, N.~Tatbul, J.~S. Turek, T.~Mattson, and A.~Muzahid,
  ``A zero-positive learning approach for diagnosing software performance
  regressions,'' in \emph{Advances in Neural Information Processing Systems
  (NeurIPS)}, 2019.

\bibitem{asselin2016anomaly}
E.~Asselin, C.~Aguilar-Melchor, and G.~Jakllari, ``Anomaly detection for web
  server log reduction: A simple yet efficient crawling based approach,'' in
  \emph{IEEE Conference on Communications and Network Security}, 2016.

\bibitem{bonneau2006cache}
J.~Bonneau and I.~Mironov, ``Cache-collision timing attacks against aes,'' in
  \emph{International Workshop on Cryptographic Hardware and Embedded Systems
  (CHES)}, 2006.

\bibitem{bossi2016system}
L.~Bossi, E.~Bertino, and S.~R. Hussain, ``A system for profiling and
  monitoring database access patterns by application programs for anomaly
  detection,'' \emph{IEEE Transactions on Software Engineering}, 2016.

\bibitem{breunig2000lof}
M.~M. Breunig, H.-P. Kriegel, R.~T. Ng, and J.~Sander, ``Lof: identifying
  density-based local outliers,'' in \emph{ACM SIGMOD International Conference
  on Management of Data (SIGKDD)}, 2000.

\bibitem{brunellaforeshadow}
M.~S. Brunella, S.~Turco, G.~Bianchi, and N.~B. Melazzi, ``Foreshadow-vmm: on
  the practical feasibility of l1 cache terminal fault attacks,'' 2018.

\bibitem{chauhan2015anomaly}
S.~Chauhan and L.~Vig, ``Anomaly detection in ecg time signals via deep long
  short-term memory networks,'' in \emph{IEEE International Conference on Data
  Science and Advanced Analytics}, 2015.

\bibitem{GRU}
K.~Cho, B.~Van~Merri{\"e}nboer, C.~Gulcehre, D.~Bahdanau, F.~Bougares,
  H.~Schwenk, and Y.~Bengio, ``Learning phrase representations using rnn
  encoder-decoder for statistical machine translation,'' \emph{arXiv preprint
  arXiv:1406.1078}, 2014.

\bibitem{das2019sok}
S.~Das, J.~Werner, M.~Antonakakis, M.~Polychronakis, and F.~Monrose, ``Sok: The
  challenges, pitfalls, and perils of using hardware performance counters for
  security,'' in \emph{IEEE Symposium on Security and Privacy (S\&P)}, 2019.

\bibitem{demme2013feasibility}
J.~Demme, M.~Maycock, J.~Schmitz, A.~Tang, A.~Waksman, S.~Sethumadhavan, and
  S.~Stolfo, ``On the feasibility of online malware detection with performance
  counters,'' \emph{ACM SIGARCH Computer Architecture News}, 2013.

\bibitem{BERT}
J.~Devlin, M.-W. Chang, K.~Lee, and K.~N. Toutanova, ``Bert: Pre-training of
  deep bidirectional transformers for language understanding,'' \emph{arXiv
  preprint arXiv:1810.04805}, 2018.

\bibitem{du2019lifelong}
M.~Du, Z.~Chen, C.~Liu, R.~Oak, and D.~Song, ``Lifelong anomaly detection
  through unlearning,'' in \emph{ACM Conference on Computer and Communications
  Security (CCS)}, 2019.

\bibitem{du2017deeplog}
M.~Du, F.~Li, G.~Zheng, and V.~Srikumar, ``Deeplog: Anomaly detection and
  diagnosis from system logs through deep learning,'' in \emph{ACM Conference
  on Computer and Communications Security (CCS)}, 2017.

\bibitem{garg2020abc}
S.~Garg, K.~Kaur, S.~Batra, G.~S. Aujla, G.~Morgan, N.~Kumar, A.~Y. Zomaya, and
  R.~Ranjan, ``En-abc: An ensemble artificial bee colony based anomaly
  detection scheme for cloud environment,'' \emph{Journal of Parallel and
  Distributed Computing}, 2020.

\bibitem{gruss2016flush}
D.~Gruss, C.~Maurice, K.~Wagner, and S.~Mangard, ``Flush+flush: a fast and
  stealthy cache attack,'' in \emph{International Conference on Detection of
  Intrusions and Malware, and Vulnerability Assessment}, 2016.

\bibitem{gruss2015cache}
D.~Gruss, R.~Spreitzer, and S.~Mangard, ``Cache template attacks: Automating
  attacks on inclusive last-level caches,'' in \emph{USENIX Security
  Symposium}, 2015.

\bibitem{gullasch2011cache}
D.~Gullasch, E.~Bangerter, and S.~Krenn, ``Cache games--bringing access-based
  cache attacks on aes to practice,'' in \emph{IEEE Symposium on Security and
  Privacy (S\&P)}, 2011.

\bibitem{guo2020specusym}
S.~Guo, Y.~Chen, P.~Li, Y.~Cheng, H.~Wang, M.~Wu, and Z.~Zuo, ``Specusym:
  Speculative symbolic execution for cache timing leak detection,'' in
  \emph{International Conference on Software Engineering}, 2020.

\bibitem{he2021new}
Z.~He, G.~Hu, and R.~B. Lee, ``New models for understanding and reasoning about
  speculative execution attacks,'' in \emph{IEEE International Symposium on
  High-Performance Computer Architecture (HPCA)}, 2021.

\bibitem{he2017secure}
Z.~He and R.~B. Lee, ``How secure is your cache against side-channel attacks?''
  in \emph{Annual IEEE/ACM International Symposium on Microarchitecture}, 2017.

\bibitem{he2019power}
Z.~He, A.~Raghavan, G.~Hu, S.~Chai, and R.~Lee, ``Power-grid controller anomaly
  detection with enhanced temporal deep learning,'' in \emph{IEEE International
  Conference On Trust, Security And Privacy In Computing (TrustCom)}, 2019.

\bibitem{he2019sensitive}
Z.~He, T.~Zhang, and R.~Lee, ``Sensitive-sample fingerprinting of deep neural
  networks,'' in \emph{IEEE Conference on Computer Vision and Pattern
  Recognition (CVPR)}, 2019.

\bibitem{he2019model}
Z.~He, T.~Zhang, and R.~B. Lee, ``Model inversion attacks against collaborative
  inference,'' in \emph{Annual Computer Security Applications Conference
  (ACSAC)}, 2019.

\bibitem{he2020attacking}
------, ``Attacking and protecting data privacy in edge-cloud collaborative
  inference systems,'' \emph{IEEE Internet of Things Journal}, 2020.

\bibitem{henning2006spec}
J.~L. Henning, ``Spec cpu2006 benchmark descriptions,'' \emph{ACM SIGARCH
  Computer Architecture News}, 2006.

\bibitem{hinton2002stochastic}
G.~E. Hinton and S.~Roweis, ``Stochastic neighbor embedding,'' \emph{Advances
  in Neural Information Processing Systems (NeurIPS)}, 2002.

\bibitem{hotelling1933analysis}
H.~Hotelling, ``Analysis of a complex of statistical variables into principal
  components.'' \emph{Journal of educational psychology}.

\bibitem{hu2021smartphone}
G.~Hu, Z.~He, and R.~B. Lee, ``Smartphone impostor detection with behavioral
  data privacy and minimalist hardware support,'' in \emph{TinyML Symposium},
  2021.

\bibitem{irazoqui2015s}
G.~Irazoqui, T.~Eisenbarth, and B.~Sunar, ``S \$ a: A shared cache attack that
  works across cores and defies vm sandboxing--and its application to aes,'' in
  \emph{IEEE Symposium on Security and Privacy (S\&P)}, 2015.

\bibitem{kayaalp2016high}
M.~Kayaalp, N.~Abu-Ghazaleh, D.~Ponomarev, and A.~Jaleel, ``A high-resolution
  side-channel attack on last-level cache,'' in \emph{Design Automation
  Conference (DAC)}, 2016.

\bibitem{khan2012workload}
A.~Khan, X.~Yan, S.~Tao, and N.~Anerousis, ``Workload characterization and
  prediction in the cloud: A multiple time series approach,'' in \emph{IEEE
  Network Operations and Management Symposium}, 2012.

\bibitem{kiriansky2018speculative}
V.~Kiriansky and C.~Waldspurger, ``Speculative buffer overflows: Attacks and
  defenses,'' \emph{arXiv preprint arXiv:1807.03757}, 2018.

\bibitem{kocher2019spectre}
P.~Kocher, J.~Horn, A.~Fogh, D.~Genkin, D.~Gruss, W.~Haas, M.~Hamburg, M.~Lipp,
  S.~Mangard, T.~Prescher \emph{et~al.}, ``Spectre attacks: Exploiting
  speculative execution,'' in \emph{IEEE Symposium on Security and Privacy
  (S\&P)}, 2019.

\bibitem{koruyeh2018spectre}
E.~M. Koruyeh, K.~N. Khasawneh, C.~Song, and N.~Abu-Ghazaleh, ``Spectre
  returns! speculation attacks using the return stack buffer,'' in \emph{USENIX
  Workshop on Offensive Technologies}, 2018.

\bibitem{lipp2018meltdown}
M.~Lipp, M.~Schwarz, D.~Gruss, T.~Prescher, W.~Haas, A.~Fogh, J.~Horn,
  S.~Mangard, P.~Kocher, D.~Genkin \emph{et~al.}, ``Meltdown: Reading kernel
  memory from user space,'' in \emph{USENIX Security Symposium}, 2018.

\bibitem{liu2015last}
F.~Liu, Y.~Yarom, Q.~Ge, G.~Heiser, and R.~B. Lee, ``Last-level cache
  side-channel attacks are practical,'' in \emph{IEEE Symposium on Security and
  Privacy (S\&P)}, 2015.

\bibitem{liu2008isolation}
F.~T. Liu, K.~M. Ting, and Z.-H. Zhou, ``Isolation forest,'' in \emph{IEEE
  International Conference on Data Mining (ICDM)}, 2008.

\bibitem{liu2016anomaly}
J.~Liu, S.~Chen, Z.~Zhou, and T.~Wu, ``An anomaly detection algorithm of cloud
  platform based on self-organizing maps,'' \emph{Mathematical Problems in
  Engineering}, 2016.

\bibitem{malhotra2015long}
P.~Malhotra, L.~Vig, G.~Shroff, and P.~Agarwal, ``Long short term memory
  networks for anomaly detection in time series,'' in \emph{European Symposium
  on Artificial Neural Networks, Computational Intelligence and Machine
  Learning}, 2015.

\bibitem{mishra2010towards}
A.~K. Mishra, J.~L. Hellerstein, W.~Cirne, and C.~R. Das, ``Towards
  characterizing cloud backend workloads: insights from google compute
  clusters,'' \emph{ACM SIGMETRICS Performance Evaluation Review}, 2010.

\bibitem{osvik2006cache}
D.~A. Osvik, A.~Shamir, and E.~Tromer, ``Cache attacks and countermeasures: the
  case of aes,'' in \emph{Cryptographers’ Track at the RSA conference}, 2006.

\bibitem{ozsoy2016hardware}
M.~Ozsoy, K.~N. Khasawneh, C.~Donovick, I.~Gorelik, N.~Abu-Ghazaleh, and
  D.~Ponomarev, ``Hardware-based malware detection using low-level
  architectural features,'' \emph{IEEE Transactions on Computers}, 2016.

\bibitem{pannu2012aad}
H.~S. Pannu, J.~Liu, and S.~Fu, ``Aad: Adaptive anomaly detection system for
  cloud computing infrastructures,'' in \emph{IEEE Symposium on Reliable
  Distributed Systems}, 2012.

\bibitem{patel2017analyzing}
N.~Patel, A.~Sasan, and H.~Homayoun, ``Analyzing hardware based malware
  detectors,'' in \emph{Design Automation Conference (DAC)}, 2017.

\bibitem{qiu2020adversarial}
H.~Qiu, T.~Dong, T.~Zhang, J.~Lu, G.~Memmi, and M.~Qiu, ``Adversarial attacks
  against network intrusion detection in iot systems,'' \emph{IEEE Internet of
  Things Journal}, 2020.

\bibitem{scholkopf2000support}
B.~Sch{\"o}lkopf, R.~C. Williamson, A.~J. Smola, J.~Shawe-Taylor, and J.~C.
  Platt, ``Support vector method for novelty detection,'' in \emph{Advances in
  Neural Information Processing Systems (NeurIPS)}, 2000.

\bibitem{vallis2014novel}
O.~Vallis, J.~Hochenbaum, and A.~Kejariwal, ``A novel technique for long-term
  anomaly detection in the cloud,'' in \emph{USENIX Workshop on Hot Topics in
  Cloud Computing}, 2014.

\bibitem{wang2020kleespectre}
G.~Wang, S.~Chattopadhyay, A.~K. Biswas, T.~Mitra, and A.~Roychoudhury,
  ``Kleespectre: Detecting information leakage through speculative cache
  attacks via symbolic execution,'' \emph{ACM Transactions on Software
  Engineering and Methodology}, 2020.

\bibitem{wang2008sigfree}
X.~Wang, C.-C. Pan, P.~Liu, and S.~Zhu, ``Sigfree: A signature-free buffer
  overflow attack blocker,'' \emph{IEEE Transactions on Dependable and Secure
  Computing}, 2008.

\bibitem{wang2014detecting}
X.~Wang and R.~Karri, ``Detecting kernel control-flow modifying rootkits,'' in
  \emph{Network Science and Cybersecurity}, 2014.

\bibitem{wang2015confirm}
X.~Wang, C.~Konstantinou, M.~Maniatakos, and R.~Karri, ``Confirm: Detecting
  firmware modifications in embedded systems using hardware performance
  counters,'' in \emph{International Conference on Computer-Aided Design
  (ICCAD)}, 2015.

\bibitem{weisse2018foreshadow}
O.~Weisse, J.~Van~Bulck, M.~Minkin, D.~Genkin, B.~Kasikci, F.~Piessens,
  M.~Silberstein, R.~Strackx, T.~F. Wenisch, and Y.~Yarom, ``Foreshadow-ng:
  Breaking the virtual memory abstraction with transient out-of-order
  execution,'' Tech. Rep., 2018.

\bibitem{yao2020miss}
Q.~Yao, Z.~He, H.~Han, and S.~K. Zhou, ``Miss the point: Targeted adversarial
  attack on multiple landmark detection,'' in \emph{International Conference on
  Medical Image Computing and Computer-Assisted Intervention (MICCAI)}, 2020.

\bibitem{yarom2014flush}
Y.~Yarom and K.~Falkner, ``Flush+reload: a high resolution, low noise, l3 cache
  side-channel attack,'' in \emph{USENIX Security Symposium}, 2014.

\bibitem{yin2017deep}
C.~Yin, Y.~Zhu, J.~Fei, and X.~He, ``A deep learning approach for intrusion
  detection using recurrent neural networks,'' \emph{IEEE Access}, 2017.

\bibitem{zhang2016cloudradar}
T.~Zhang, Y.~Zhang, and R.~B. Lee, ``Cloudradar: A real-time side-channel
  attack detection system in clouds,'' in \emph{International Symposium on
  Research in Attacks, Intrusions, and Defenses (RAID)}, 2016.

\bibitem{zhang2014cross}
Y.~Zhang, A.~Juels, M.~K. Reiter, and T.~Ristenpart, ``Cross-tenant
  side-channel attacks in paas clouds,'' in \emph{ACM Conference on Computer
  and Communications Security (CCS)}, 2014.

\bibitem{zhou2018hardware}
B.~Zhou, A.~Gupta, R.~Jahanshahi, M.~Egele, and A.~Joshi, ``Hardware
  performance counters can detect malware: Myth or fact?'' in \emph{Asia
  Conference on Computer and Communications Security (AsiaCCS)}, 2018.

\end{thebibliography}
%

\newpage
\appendix


\begin{table}[h]
\centering
\caption{List of all hardware performance counters.}\label{tab:cs:hpc-list}
\resizebox{\linewidth}{!}{
\begin{tabular}{|l|l|}
\hline
\textbf{HPC} & \textbf{Description}                                                                  \\ \hline
Instruction                     & Number of instructions                                             \\ \hline
Load                            & Number of memory loads                                             \\ \hline
Store                           & Number of memory stores                                            \\ \hline
L1D read miss                   & Number of L1 data cache read misses                                \\ \hline
L1D write miss                  & Number of L1 data cache write misses                               \\ \hline
L1D prefetch miss               & Number of L1 data cache prefetch misses                            \\ \hline
L1I read miss                   & Number of L1 instruction cache read misses                         \\ \hline
LLC read access                 & Number of Last level cache read accesses                           \\ \hline
LLC read miss                   & Number of Last level cache read misses                             \\ \hline
LLC write access                & Number of Last level cache write access                            \\ \hline
LLC write miss                  & Number of Last level cache write misses                            \\ \hline
LLC prefetch access             & Number of Last level cache prefetch accesses                       \\ \hline
LLC prefetch miss               & Number of Last level cache prefetch misses                         \\ \hline
DTLB read access                & Number of data translation lookaside buffer read accesses          \\ \hline
DTLB read miss                  & Number of data translation lookaside buffer read misses            \\ \hline
DTLB write access               & Number of data translation lookaside buffer write accesses         \\ \hline
DTLB write miss                 & Number of data translation lookaside buffer write misses           \\ \hline
ITLB read access                & Number of instruction translation lookaside buffer read accesses   \\ \hline
ITLB read miss                  & Number of instruction translation lookaside buffer read misses     \\ \hline
BPU read access                 & Number of branch prediction unit read accesses                     \\ \hline
BPU read miss                   & Number of branch prediction unit read misses                       \\ \hline
Cache node read access          & Number of cache node read accesses                                 \\ \hline
Cache node read miss            & Number of cache node read misses                                   \\ \hline
Cache node write access         & Number of cache node write accesses                                \\ \hline
Cache node write miss           & Number of cache node write misses                                  \\ \hline
Cache node prefetch access      & Number of cache node prefetch accesses                             \\ \hline
Cache node prefetch miss        & Number of cache node prefetch misses                               \\ \hline
Cycles                          & Number of cycles                                                   \\ \hline
Branch instructions             & Number of branch instructions                                      \\ \hline
Branch prediction miss          & Number of branch prediction misses                                 \\ \hline
Page faults                     & Number of page faults                                              \\ \hline
Context switch                  & Number of context switches                                         \\ \hline
Stall during issue              & Number of stalled cycles during instruction issue                  \\ \hline
Stall during retirement         & Number of stalled cycles during instruction retirement             \\ \hline
\end{tabular}
}
\end{table}

\newcommand{\stallduringissue}{\begin{tabular}[c]{@{}c@{}}\textbf{Stall during}\\ \textbf{issue}\end{tabular}}
\newcommand{\stallduringretirement}{\begin{tabular}[c]{@{}c@{}}\textbf{Stall during}\\ \textbf{retirement}\end{tabular}}
\newcommand{\dreadmiss}{\begin{tabular}[c]{@{}c@{}}L1D read\\ miss\end{tabular}}
\newcommand{\ireadmiss}{\begin{tabular}[c]{@{}c@{}}L1I read\\ miss\end{tabular}}

\begin{table}[h]
\caption{Top-10 important events for each cloud benchmark. Bold means the event ranks top 10 for all benchmarks.} \label{tab:cs:feature-selection-per-task}
\centering
\resizebox{\linewidth}{!}{
\begin{tabular}{|c|c|c|c|c|c|}
\hline
Rank & \begin{tabular}[c]{@{}c@{}}ML Training\\ (Pytorch)\end{tabular} & \begin{tabular}[c]{@{}c@{}}Stream Server\\ (FFserver)\end{tabular}
              & \begin{tabular}[c]{@{}c@{}}Database\\ (Mysql)\end{tabular}      & \begin{tabular}[c]{@{}c@{}}Web Server\\ (Nginx)\end{tabular}
              & MapReduce                                                     \\ \hline
1    & \textbf{Instruction}                 & \textbf{Cycles}                    & \textbf{Cycles}
              &  \stallduringissue          & \textbf{Instruction}               \\ \hline
2    & \textbf{Load}                        & \stallduringissue                 & \stallduringissue
              & \stallduringretirement      & \stallduringissue       \\ \hline
3    & \stallduringretirement     & \stallduringretirement   & \stallduringretirement
              & \textbf{Cycles}                      & \textbf{Cycles}                    \\ \hline
4    & Store                                & \textbf{Instruction}               & \textbf{Instruction}
              & \textbf{Load}                        & \stallduringretirement   \\ \hline
5    & \stallduringissue         & \textbf{Load}                      & \textbf{Load}
              & \textbf{DTLB read}                   & \textbf{Load}                      \\ \hline
6    & \textbf{DTLB read}                   & \textbf{BPU read}                  & \textbf{BPU read}
              & Branch                                      & \textbf{DTLB read}                 \\ \hline
7    & \textbf{BPU read}                    & \textbf{DTLB read}                 & \textbf{DTLB read}
              & \ireadmiss                               & Branch                                    \\ \hline
8    & DTLB write                                  & \textbf{Store}              & Store
              & \textbf{Instruction}                 & \textbf{BPU read}                  \\ \hline
9    & \textbf{Cycles}                      & DTLB write                                & DTLB write
              & \textbf{BPU read}                    & DTLB write                                \\ \hline
10   & \dreadmiss                               & \ireadmiss                             & \ireadmiss
              & Context Switch                              & \textbf{Store}                     \\ \hline
\end{tabular}
}
\end{table}

\end{document}